\documentclass[12pt,preprint]{aastex}

\usepackage{graphicx}
\usepackage{epstopdf}

\def\m{$\mu$m}
\def\ab{$\sim$}
\def\p{$\pm$}

\def\Lya{Ly$\alpha$ }

\def\ab{$\sim$}
\def\deg{\ifmmode ^{\circ}
         \else $^{\circ}$\fi} 

\shorttitle{PAH Emission Within Lyman Alpha Blobs}
\shortauthors{Colbert et al.}

\begin{document}
\title{PAH Emission Within Lyman Alpha Blobs}
\author{James W. Colbert\altaffilmark{1}, Claudia
  Scarlata\altaffilmark{1}, Harry Teplitz\altaffilmark{1},
Paul Francis\altaffilmark{2}, Povilas Palunas\altaffilmark{3}, Gerard 
M. Williger\altaffilmark{4}, Bruce Woodgate\altaffilmark{5}}

\altaffiltext{1}{Spitzer Science Center, California Institute of Technology,
    Pasadena, CA 91125}

\altaffiltext{2}{Research School of Astronomy and Astrophysics, The Australian National University, 
    Canberra, ACT 0200, Australia}

\altaffiltext{3}{Las Campanas Observatory, La Serena, Chile}

\altaffiltext{4}{Dept. of Physics \& Astronomy, University of Louisville, Louisville, KY 40292}

\altaffiltext{6}{NASA Goddard Space Flight Center, Greenbelt, MD 20771}

\begin{abstract}

We present {\it Spitzer} observations of \Lya Blobs (LAB) at
z=2.38--3.09. 
The mid-infrared ratios (4.5/8\m\ and 8/24\m ) indicate that \ab 60\% of LAB
infrared counterparts are cool, consistent with their infrared
output being dominated by star formation and not active galactic
nuclei (AGN).  The rest have a
substantial hot dust component that one would expect from an AGN or an extreme
starburst. Comparing the mid-infrared to submillimeter fluxes (\ab
850\m\ or rest frame far infrared) also indicates a large percentage (\ab 2/3) of the
LAB counterparts have total bolometric energy output dominated by star
formation, although the number of sources with sub-mm detections or meaningful upper 
limits remains small (\ab 10). We obtained Infrared Spectrograph (IRS) spectra of 6 infrared-bright
sources associated with LABs. Four of these sources have
measurable polycyclic aromatic hydrocarbon (PAH) emission features, indicative of significant star
formation, while the remaining two show a featureless continuum,
indicative of infrared energy output completely dominated by an
AGN. Two of the counterparts with PAHs are mixed sources, with PAH
line-to-continuum ratios and PAH equivalent widths indicative of
large energy contributions from both star formation and AGN. 
Most of the LAB infrared counterparts have large stellar masses, around 10$^{11}$
M$_{\odot }$. There is a weak trend of mass upper limit with the \Lya luminosity of the
host blob, particularly after the most likely AGN contaminants are
removed. The range in likely energy sources for the LABs found in this
and previous studies suggests that there is no single source of power
that is producing all the known LABs.

\end{abstract}


\section{Introduction}

The Ly$\alpha$ blob (LAB) remains one of the great mysteries of the high
redshift universe.  While these extended \Lya nebulae are similar in 
extent (5-20$\arcsec$ or $\sim$50-150 kpc) and \Lya flux (\ab 10$^{43-44}$ ergs s$^{-1}$) to 
high-redshift radio galaxies, blobs are radio quiet and are therefore unlikely to arise 
from interaction with jets. They are found almost exclusively within high-redshift galaxy 
over-densities \citep{mat09,pre08,pal04,ste00} with none found so far at even moderate
redshift \citep[z$<$0.8;][]{kee09}, suggesting strong evolution. After more than a
decade of searching, there are still only a
handful of the truly giant ($>$50 arcsec$^2$, $>$5$\times$10$^{43}$ ergs
s$^{-1}$) LABs known. However, deep searches \citep{mat04,sai06}
show that the blobs are part of a continuous size distribution of 
resolved \Lya emitters.

With most lying at the density peak of high
redshift structures \citep{mat09,mat04,pal04}
and with number densities comparable to galaxy clusters in the nearby and
high-z universe \citep[10$^{-5}$--10$^{-6}$ Mpc$^{-3}$;][]{yan09}, it
seems likely that the giant LABs are at the very least signposts
for regions of massive galaxy assembly, if not the progenitors of the
massive elliptical galaxies themselves. The limited
{\it HST} imaging of these objects to date shows some evidence for
interaction and merger of multiple compact objects
\citep{cha04,fra01}. Major mergers of high-mass 
galaxies, like those predicted to build giant ellipticals, must be
occurring at higher redshift \citep[i.e.,][]{nar10,ste09}. 
The LAB may represent an opportunity for the study of the merger
formation of the most massive galaxies.    

One of the biggest unknowns of the blobs is the source of their energy.
\cite{mat04} found that for at least a third of LABs, including all
the biggest and brightest of them, the galaxies within the blob do not
emit enough rest-wavelength UV light to excite such a vast quantity of
hydrogen gas. The exciting ultraviolet illumination could be escaping along different 
lines of sight from an obscured AGN \citep{bas04}. Many
LABs contain luminous X-ray counterparts \citep{gea09,yan09}, while
others show powerful AGN emission lines \citep{sca09,dey05,pas96}. Alternatively,
outflows could be driving great plumes of gas into the surrounding
ambient medium and producing shocks. While AGNs are known to drive
some outflows, supernova-driven superwinds are also a viable model
\citep{tan00,ohy04}. Such outflows would likely take the form of
immense bubbles or shells expanding outwards from the galaxy, some
evidence of which has been claimed in the brightest LABs
\citep{mor07}. 

Integral field spectroscopy of the LABs show the large \Lya velocty
widths and structure consistent with superwind outflows
\citep{bow04,wil05,wei10}, although the
systems are complicated enough that other velocity models, like rotation, can
not be completely ruled out. Cooling flows have also been 
suggested as a possible source of the extended \Lya emission \citep{hai00,fra01,dij06}.  The
primary evidence to support the cooling model are LABs with no apparent
internal power source, even out to the mid-infrared
\citep{nil06,smi08}, and sources with He II emission with weak to no 
measurable CIV emission \citep{pre09,sca09}, indicative of the lower
temperature gas emission that one would expect from a cooling flow \citep{ber10}. 

Mid-infrared and submillimeter imaging show that it is very common
that the extended nebulae of LABs contain sources of extreme infrared
luminosity. Powerful {\it Spitzer} 24\m\ sources have been found within
roughly ten LABs, and almost all the most luminous ones, \citep{web09,gea07,col06b,dey05}, with fluxes
of 0.05-0.86 mJy. Submillimeter flux has been measured for a similar
number \citep{cha04,sma03,gea05,bee08}.
Mid-infrared colors suggest significant quantities of hot dust for
these infrared-bright sources \citep{web09}, although whether
from AGN or extreme star formation remains unclear.  

In this paper we discuss mid-infrared and submillimeter
  observations of mid-infrared sources identified within LABs from four different fields, with
  $z$=2.38-3.09.  We examine both their  {\it Spitzer} mid-infrared
    flux ratios and mid-infrared to sub-mm flux ratios and compare
    them to models in order to identify their likely power source, AGN
    or star formation. We then look at {\it Spitzer} IRS spectra of 6
     sources for an even more definitive AGN/star formation
     separation. Finally, we look at the possible masses of these
     infrared sources from within the LABs. We assume an $\Omega _{M}$=0.3, 
$\Omega _{\Lambda }$=0.7 universe with H$_o$=70 km s$^{-1}$ Mpc$^{-1}$.

\section{Data Acquisition and Reduction}

\subsection{LAB Nomenclature}

Because of their rarity, most LABs are discovered in small numbers,
often one or two at a time. This has led to a naming system where the
LABs in question are often just referred to as Blob 1 or Blob 2, if
they are given any specific name at all. That creates a problem for a
work such as this one, where multiple LABs are being discussed from
multiple fields, all known as Blob 1 or LAB1 or B1 or something
similar. Since LABs are large, they occasionally have multiple counterparts
associated with them, which can make it even more difficult to identify the
correct object under discussion.

To address this issue we will use the following naming system in this
paper. All LABs will be referred to as:

LAB[number][letter]\_J[coordinate of associated field center]

The {\it [number]} used is that which has been associated with the
LAB in previous publications, if there has been any. If no
previous number has been assigned we will begin the labeling at LAB1
and count upwards from there. In many cases, the original numbering
included all the detected \Lya emitting sources in the field so just
because there is a LAB6 does not indicate there must be a LAB5. While
these "missing'' numbers might create some mild confusion, we found
this system to be superior to the certain confusion that would result
from changing the numbers that have already been used to identify
these sources from one paper to the next.   

The {\it [letter]} refers to the counterpart of the LAB being
discussed. Most blob counterparts have only one associated counterpart and will 
therefore be given no letter identifier, so the existence of a letter
implies that at some point in the literature, multiple
components have been assigned to the LAB. Discussions of the LABs themselves will never be given a letter.

The {\it [coordinate of the associated field center]} is an eight
digit code giving the right ascension and declination (J2000) of the field
with which the LAB is associated. Most of these fields have been
observed multiple times with slightly different centers, but for this
paper we have chosen J2217+0017, J1714+5014, J1434+3317, and
J2143-4423 to represent the SSA22, 53w002, NDWFS, and J2143-4423
fields in this study (see below). We found this method of only providing a field
coordinate superior to a full 12-14 digit coordinate designation as it makes the
discussion of object names significantly less cumbersome and keeps
a clear connection with the field and/or structure they have been
found within. 

Table 1 provides the list of all the LAB infrared counterparts
included in this paper, providing both the new name as well as any previous
names by which the LABs have been known. 

\subsection{IRAC and MIPS data}

The LABs presented in this paper are spread across
four different fields: J2143-4423, SSA22, 53w002, and the NOAO Deep
Wide-Field Survey. We assembled all available {\it Spitzer}
mid-infrared imaging data, which includes both IRAC and MIPS imaging.

The data for the J2143-4423 field (LAB1\_J2143-4423, LAB6\_J2143-4423,
LAB7\_J2143-4423) come from GO-3699 (PI: Colbert) and were done in a 3 $\times$ 5 raster map covering
15$\arcmin$ $\times$ 25$\arcmin$, centered at $\alpha$= 21$^h$42$^m$35$^s$,
$\delta$ = -44$\deg$27$\arcmin$ (J2000.0). The total integration times per pixel
were 1800 s for IRAC channels 1-4 and 1818 s for MIPS 24\m . This
reached 3 $\sigma$ depths of 1, 7, and 40 $\mu$Jy for 4.5, 8, and 24
\m\ respectively. 

For the SSA22 field \citep[see][]{ste00,mat04} we
examine the 16 LABs with both an isophotal area greater than 20 square
arcseconds and submillimeter data coverage, excluding only LAB4\_J2217+0017 which
lies too close to a bright object for an uncontaminated analysis. 
We assembled the {\it Spitzer} imaging from multiple programs taken at
different epochs, centered roughly at $\alpha$= 22$^h$17$^m$40$^s$,
$\delta$ = +00$\deg$17$\arcmin$ (J2000.0). The IRAC and MIPS data come
from four programs: GO 30600 (PI: Colbert), GO 3473 (PI: Blain), and GTO 64 \& GTO 30328
(both PI: Fazio). The IRAC data was previously presented in
\cite{web09}, but we re-extracted the photometry to ensure uniformity among
our several fields and also extracted to slightly deeper (3$\sigma$)
depths. The MIPS data presented here includes new data (GO 30600, 
see below) not presented in \cite{web09}, covering more sources 
and greatly increasing the 24 \m\ depth.  
 
The SSA22 IRAC data from GTO 64 was a deep pair of single 
pointings covering most of the known SSA22 blobs, GTO
30328 was a 3 $\times$ 4 raster map, while GO 3473 was a smaller
2 $\times$ 3 raster map that lies to the east of the majority of the
known LAB positions. Altogether the combined IRAC image maps out a
region roughly 20 $\times$ 26 arcminutes, ranging in depth from 1500 to
15,000 seconds. Most of the LABs lie at depths of 4000 s or
greater, roughly corresponding to 3 $\sigma$ depths of 0.5 and 5
$\mu$Jy at 4.5 and 8\m\ respectively. Only one of the sub-mm detected
blobs (LAB10\_J2217+0017) completely falls off the area covered by IRAC.

The SSA22 MIPS data from GTO 64 is a single pointing (1120 s), GTO
30328 is a 3 $\times$ 3 map built from a set of cluster offsets
(1200 s per pixel), and GO 3473 was a set of pointings targeting known
sub-mm galaxies (1200 s). GO 30600 consisted of two parts. The
first was a set of three pointings targeting the sub-mm detected blobs
on the outskirts of the previously taken 24\m\ data (2700 s each). The
second part was an extremely deep pointing on the central portion of the
field (10,800 s) where the biggest LABs \citep[LAB1\_J2217+0017 \&
LAB2\_J2217+0017;][]{ste00} are located. Assembled altogether the MIPS data covers an
area roughly 15 $\times$ 26 arcminutes, with the regions covering
the LAB locations ranging from 1200 - 13,000 seconds, with all sub-mm
detected LABs observed for at least 3300 seconds, which provides
a 3 $\sigma$ depth of \ab 30 $\mu$Jy. 

The 53w002 field, centered at $\alpha$ = 17$^h$14$^m$20$^s$, $\delta$
= +50$\deg$15$\arcmin$ (J2000.0), contains two LABs \citep{kee99}. The IRAC data
come from GTO 211 (PI: Fazio) and were obtained in a single pointing of 3300 s
in all four IRAC bands. The MIPS data are a combination of GO 3329 (PI:
Stern), a single 500s pointing, and GO 20253 (PI: Im), a much deeper 
exposure of 18720 seconds. The 3 $\sigma$ depths obtained are 0.6, 5, and
12 $\mu$Jy for 4.5, 8, and 24 \m\ respectively, not accounting for the effect
of confusion noise which becomes significant in such a deep 24 \m\
pointing. Both blobs in the field have mid-infrared counterparts that are bright
enough that confusion is not an issue.

The NOAO Deep Wide-Field Survey (NDWFS) contains one reported bright
blob \citep{dey05}, located at $\alpha$ = 14$^h$34$^m$11.0$^s$, $\delta$
= +33$\deg$17$\arcmin$31$\arcsec$ (J2000.0). We do not attempt to
re-reduce or extract new photometry for this source and use the IRAC
and MIPS fluxes already reported in \cite{dey05}.

Both the IRAC and MIPS data were combined using the MOPEX package
available from the {\it Spitzer} Science Center (SSC). For the IRAC data we 
started with the artifact-mitigated (.cbcd) files where available (GTO 
30328). For the programs that had not been yet been reprocessed above
pipeline S16 (the first .cbcd files begin at S17, released Jan. 08), we used
the basic bcd and applied the IRAC Artifact Mitigation code (available from the
SSC) ourselves. For our extragalactic fields, which contain very few
bright stars or extended sources, we saw no significant difference
between the two methods. We then removed a residual offset from each
frame by subtracting the median of each frame itself (i.e., removing a
constant sky), medianing all the subtracted frames, and then
subtracting that resulting offset frame from all the original frames. This
prevents a gradient from appearing in the final images. This is the
data product we place into MOPEX to produce the final, combined image. 

For the MIPS data, a median flat was created from all available data taken near in time,
normalized, and divided into each image. After the Overlap Correction
module was executed by MOPEX we found that we still had some trouble 
getting all the various sky levels to match up, so an additional sky
constant was also subtracted before final combination of all the MIPS
frames.   
 
\subsection{Optical and Near-Infrared Data}

The optical and NIR data for the J2143-4423 field (U, B, G, R, J and
H) were acquired at the Cerro Tololo Inter-American Observatory (CTIO) 
4 meter telescope, using the Mosaic II imager, and the near Infrared Side Port Imager 
camera (ISPI). The seeing ranged from 0.9--1.4$\arcsec$, reaching
depths of \ab 26 for U and B, \ab 24 in R, and \ab 23 in J and H. For
further details see \cite{sca09}. All magnitudes are AB magnitudes here and
throughout the paper.

\subsection{IRS Spectra}

As part of the {\it Spitzer} GO 30600, we targeted all known bright
($>$ 200 $\mu$Jy) 24$\mu$m sources associated with LABs,
including the those within the J2143-4423 110 Mpc filament structure at
$z$=2.38 \citep{pal04,col06b} and one within the
possible proto-galaxy structure around radio galaxy
53w002 at $z$=2.39 \citep{pas96}.  Altogether
we targeted five mid-infrared sources associated with four \Lya
blobs. LAB6\_J2143-4423 contains two powerful MIPS
24\m\ sources \citep{col06b}, both of which have been confirmed
to be at the redshift of the blob \citep{sca09}. At the time of observation there
was one more known mid-infrared bright (0.86 mJy at 24\m\ ) LAB1\_J1434+3317
discovered in the NOAO Deep Wide-Field Survey by \cite{dey05} for which IRS spectra had
already been taken. We include its IRS data as part of our analysis. The complete list with
coordinates is presented in Table 2.

There are no $>$ 200 $\mu$Jy 24$\mu$m-emitting LABs in the most well studied
LAB field of all, the $z$=3.09 SSA22 field \citep{ste00}, despite
many of them having significant sub-mm detections. This is likely the result of the strong
7.7\m\ feature having shifted out of the MIPS 24\m\ filter. For these sources
we only examine their mid-IR to sub-mm ratios.

The mid-infrared spectra were all taken with the {\it Spitzer} Infrared Spectrograph (IRS) using
the 1st order of the Long-Low module (LL1), which is sensitive from 19.5 - 38.0 \m . The LL1 module 
has a spatial resolution of 5.1/arcsec per pixel and  a wavelength resolution of R=58-116, with the
$\Delta \lambda$ an approximately constant 0.17\m\ per pixel. The aperture of the LL1 long slit is 
10.7$\times$ 168 arcseconds.

The IRS data were acquired from the {\it Spitzer} cycle 3 GO 30600 program, mostly taken June 13-21, 2007,
although one spectrum (LAB18\_J1714+5015) was taken September 16, 2006. These spectra were acquired using the 
IRS Spectral Mapping Astronomical Observation Template (AOT), placing the source at six separate 
positions along the length of the slit, each separated by 20\arcsec . A high accuracy peak-up 
observation on a nearby 2MASS star was done for each observation. We used the 120 second ramp
exposure throughout, producing total exposure times per source of 84 -- 528 minutes. 

We began our data reduction with the Basic Calibrated Data (BCDs) produced by the Spitzer pipeline 
S16.1. A significant latent charge builds up on the detector over long observations that varies 
depending on wavelength. To correct this we measure a median background value for each wavelength
row and then fit a slope to the change in background with time. Using the slope at each wavelength 
we derive the corresponding latent build-up for each frame and subtracted it.

We then averaged together all the exposures taken at the same position within each Astronomical
Observation Request (AOR), producing six two-dimensional spectra per AOR. We also produced a separate 
sky frame for each position using all the images taken at the other five positions medianed together. 
We subtracted each sky frame from the its corresponding image and then ran the program IRSCLEAN, 
provided by the {\it Spitzer} Science Center (SSC), in order to remove all remaining rogue pixels or 
cosmic ray strikes. We then extracted the one-dimensional spectra for
all six positions using the optimal extraction option within the SSC 
tool SPICE (v.2.1.2). We used a smaller than standard aperture (5 
pixels at 27\m , width scaling with wavelength), in order to avoid
flux from two nearby bright sources and to reduce the inclusion of noise from blank sky. This non-standard aperture alters the SPICE 
generated calibration, but we use the known 24\m\ fluxes for final calibration so this was not an issue. 

The initial sky subtraction does not always produce a perfect zero background and can not account for 
contamination from brighter, nearby sources. To address this we extracted one dimensional sky spectra 
using the same method and aperture as that for the sources, only offset by 30 arcseconds. Two sky 
spectra, one from each side, are extracted, except for the spectra closest to the edge of the slit, 
where only one is extracted. We median all the sky 
spectra together and apply a 1.7\m\ wide boxcar smoothing, before subtracting this one dimensional sky 
from the extracted source spectra. This secondary sky subtraction adds \ab 2-5\%
to the noise of the the extracted source spectra, but can make significant corrections to the spectral 
shape, particularly in the two cases (LAB1\_J2143-4423 and LAB7\_J2143-4423) with bright nearby sources.          

Finally, we average the spectra from all the positions into a final spectrum using a 2 sigma clipping. 
In the case of LAB1\_J2143-4423 the data was taken in two AORs with slightly different roll angles. This minor 
difference in position angles produced a large difference in the contamination from a nearby quasar, producing 
significantly more noise (60\% higher). Therefore we applied a variance weighting to the combination 
the LAB1\_J2143-4423 spectra. In all the other cases the noise was constant across spectra and no weighting was 
required.

As part of our analysis we also include one IRS spectrum (LAB1\_J1434+3317) taken as part as part of the GTO 15 
program in February and June of 2005. This spectrum was taken using the IRS Staring AOT, which takes 
exposures at two separate positions along the slit length, separated by \ab 55 arcseconds. Because of
the smaller number of positions, instead of subtracting a two-dimensional medianed sky from each 
position we simply subtracted one nod from the other. We also note the BCDs were produced by the S15.3 
pipeline. Otherwise the data reduction is identical to those spectra taken in IRS Spectral Mapping 
mode.   

\section{Photometry}

The LABs are large and commonly cover multiple IRAC
sources, so we used the positions of the brightest IRAC 8\m\ sources
as the likely source of any 24 \m\ or 850\m\ emission. This approach
produced a good match between the centers of the associated 8 and 
24\m\ sources in all cases where both are strongly detected. For
sources with no 8\m\ detections within the blob (5 sources from
SSA22), the position of the 3.6\m\ source within the LAB is used. None
of the sources without 8\m\ detections had 24\m\ or 850\m\ detections, 
meaning they are not included in most of the following analysis of
flux density ratios. 

Two sources had multiple strong 8\m\ sources,
each with measurable 24\m\ fluxes, within the LAB: the LAB6\_J2143-4423
\citep{col06b,sca09} and LAB1\_J2217+0017\citep{gea07,web09}. For the
case of LAB6\_J2143-4423, an examination of the full SED and IRS
spectra \citep[see][and below]{sca09} reveals that the counterpart LAB6a\_J2143-4423
is powered by an AGN while LAB6b\_J2143-4423 is powered by star formation, making it
almost a certainty that the sub-mm flux identified at that position is
coming from LAB6b\_J2143-4423 and all further analysis assumes that. The two
sources in LAB1\_J2217+0017, on the other hand, have very similar mid-IR colors,
so it is not possible to discriminate from which 8\m\ source
the submillimeter might have originated. Using both OVRO and VLA 21 cm
observations, \cite{cha04} found emission near the most northern
of the two 8\m\ objects \citep[LAB1a\_J2217+0017; ][]{gea07}. However, there is 
evidence that the 850\m\ flux for this source is likely distributed across
multiple sources\citep{mat07}, indicating that the second 8\m\ source
(LAB1b\_J2217+0017) may also be a significant contributor to the total submillimeter
flux. For our analysis of 24/850\m\ we therefore combined the two
sources together (see more below).

MIPS 24\m\ photometry was derived from PRF-fitting using APEX, part of
the {\it Spitzer} MOPEX software tool \cite{mak05}. In cases of 24\m\ non-detections we extracted limits
from the data using the known 8\m\ positions. The rest of the
photometry was done using apertures. For the U, B, R, J, \& H bands, we
derived the flux densities using apertures of radius 
\ab 1.5$\times$FWHM (1.72\arcsec , 2.10\arcsec, 1.95\arcsec,
1.35\arcsec, 1.35\arcsec). For the IRAC channels 3 and
4, we used a 3.6\arcsec aperture, but for channels 1 and
2 we used a 2.4\arcsec aperture to avoid crowding issues. All the
above apertures only required moderate aperture corrections of 10-22\%.

Substantial submillimeter imaging has been published on many of the major
fields known to contain LABs, including the $z$=3.09 SSA22 field
\citep{gea05,cha04}, the $z$=2.38 J2142-4423 cluster \citep{bee08},
and the $z$=2.39 53w002 field \citep{sma03}.   

The SSA22 field was imaged at 850\m\ using the Submillimetre Common User
Bolometer Array (SCUBA) on the James Clerk Maxwell Telescope reaching
1$\sigma$ noise limits of 1.5 mJy using the pointed photometry mode
and 5.3 mJy from a more shallow scan-map. Five blobs were detected with fluxes ranging from 
4.9-16.8 mJy \citep{gea05}. All but the brightest one were found in
the deeper pointed observations. There are also Submillimeter Array \citep[SMA;
][]{mat07} and Atacama Submillimeter Telescope Experiment
\citep[ASTE; ][]{koh08} observations of the brightest blob (LAB1\_2217+0017) in the
field. 

The J2143-4423 cluster was imaged at 870\m\ (345 GHz) using the
LABoCA \citep{sir07,sir08} instrument installed on the Atacama
Pathfinder Experiment \citep[APEX,][]{gus06} reaching a 1$\sigma$
point source sensitivity of 1.4--7 mJy, depending on distance from
center of the field. LAB7\_J2143-4423 is robustly detected (8.4\p 1.0 mJy),
while LAB6\_J2143-4423 is only marginally detected (4.9 \p 2.0 mJy).

The SCUBA 850\m\ observations of the 53w002
field reach 1-1.5 mJy noise levels and detect one of the two known
blobs there (LAB18\_J1714+5015; 5.6\p 0.9 mJy). 

For ease of discussion, the rest of the paper will use 850\m\ to describe
either the 850\m\ SCUBA data or the 870\m\ LABoCA data. It is assumed
that at the level of accuracy of the reported sub-mm fluxes there are
no measurable differences between the two sub-mm bands.

We present the flux densities of all the LAB counterparts presented in
this paper in Table 1.

\section{Results}

\subsection{Spitzer Flux Density Ratios}

The mid-infrared properties of warm ULIRGs, those dominated by AGN radiation, are significantly different
than those of cool ULIRGs powered by star formation. The SEDs of warm ULIRGs rise rapidly and 
steadily through the mid-infrared, roughly obeying a power law, while the cool ULIRG SEDs do not rise
steeply until far into the mid-infrared ($>$ 10$\mu$m). 
The other significant difference is the presence of strong polycyclic aromatic hydrocarbon (PAH) 
features in cool ULIRGs, which can be compared to the continuum to determine the fraction of infrared 
luminosity generated by star formation \citep{lut98,rig99}. 

Because of these mid-infrared color differences, plots of {\it Spitzer}
mid-infrared IRAC colors can be powerful tools for identifying AGN and
separting them from starbursts at low redshifts
\citep{ste07,ste05,lac04}. By including MIPS 24\m\ one can then continue
this analysis to much higher redshift \citep[i.e.,][]{web09,pop08}. We
plot all the mid-infrared sources associated with
LABs that have 8\m\ detections in Figure 1, which is a comparison of  the
24/8\m\ flux density ratio to the 8/4.5\m\ flux density ratio. 

Both \cite{web09} and \cite{pop08} created just such a
mid-infrared color-color plot for their respective samples of LABs and submillimeter 
galaxies. The rectangular box represents the
region \cite{pop08} identified as the location of sources
starburst-dominated in the mid-infrared. The higher  8/4.5\m\ flux ratios
are generally only obtainable by AGN, as only their SEDs should be
that steep at these high redshifts (z$>$2). Submillimeter sources with similar redshifts to
the LABs in this study, taken from \cite{pop06}, have very similar 24/8 and 8/4.5\m\
flux ratios, with mean colors that are the same, to within one
standard deviation, as that of the LAB counterparts presented in Figure 1.

For comparison we over-plot a series of models derived from
\citet[hereafter CE01]{cha01}. As CE01 is based on low redshift infrared galaxies, we
applied an evolution in specific star formation to account for the
expected build-up of stellar mass with time, a factor of (1+z)$^{2}$.
This redshift evolution roughly matches the evolution in integrated specific star
formation found for the massive galaxies studied by \cite{pap06},
erring on the side of more conservative (i.e. less)
evolution. Stronger evolution in specific star formation would produce
larger  24\m\ to 8\m\ ratios, moving the star formation ends of all
the models upwards on the plot. 

The CE01 models cover a range in luminosity, so we
selected three that are representative of that entire range. Changing 
exactly which models are plotted makes no difference to the results or
discussion we present here. Finally we mixed the star forming CE01
SED models with the SED of an AGN (Mrk 231) to produce a range of 
ratios from AGN to star formation-dominated. On the plot the models
start to the left and/or high above as pure star formation and
converge on pure Mrk 231 around 8/4.5\m\ \ab 2. The 
z=2.4 models (solid lines) cover a much larger range in 24/8\m ratio
than the z=3.1 models (dotted lines) because of the possible presence
of the large 7.7\m PAH feature at that redshift.

Nearly two thirds (59\%; 10 of 17) of the LAB-associated sources fall within the star-formation
rectangle, 29\% have mid-infrared colors clearly indicative of AGN,
while the two remaining sources (12\%) are borderline, but likely also
contain significant AGN contribution. On the surface this would
indicate that the majority of sources associated with LABs are powered
by starbursts. However, even an AGN dominant in the infrared can still have significant contribution from
stars at rest wavelength 1\m\, which has the effect of flattening its slope
considerably, i.e. it decreases the 8/4.5\m\ ratio. For instance, the
Mrk 231 ratios presented here (the converging end point of the models we plot
in Figure 1) differ somewhat from the Mrk 231 model ratios
presented in \cite{pop08}, particularly at z\ab 3. This is likely a result of their
use of a Mrk 231 model based only on mid-infrared data (a \cite{rig99}
mid-infrared spectrum combined with a fit to {\it Infared Astronomical
Satellite} photometry). The Mrk 231 model we use comes from a multi-component
fit \citep{arm07} that includes the critical stellar component fit to near-infared data
points, which has a strong effect on the resulting 8/4.5\m\ ratio at
z$>$2. 

In addition, at these redshifts these ratios are extremely
sensitive to hot dust, which one would expect to be dominated by any
AGN, even if the AGN was not a significant contributor to the total
infrared luminosity. Alternatively, an extreme starburst can also
produce a large amount of emission from hot dust, which could produce 
similar IRAC colors to an AGN \citep{yun08}.
This diagnostic plot makes no direct measurement of the far infrared luminosity, the
source of which we are trying to determine.
   
\subsection{Mid-IR to Sub-mm Ratios}

At the redshifts of LABs with sub-mm imaging ($z$=2.4-3.1), 850\m\ is sensitive to
rest-wavelength 200-250\m , past the peak of  far infrared cold dust
emission, where flux density is beginning to fall
like a Rayleigh-Jeans law (F$_{\nu } \propto \nu ^{2}$). {\it Spitzer}
MIPS 24\m\ observations are sensitive to the rest wavelength 6-7\m\
mid-infrared, blueward of the far-infrared peak and very sensitive to
both hot dust and the presence of PAHs (although less so for PAHs at the higher
redshift, see discussion below). {\it Spitzer} IRAC observations at
these redshifts are not really sensitive to dust emission but instead
can be dominated by stellar emission. In particular, 8\m\ is roughly
equivalent to rest wavelength K-band, an excellent tracer of total
stellar mass.

The combination of {\it Spitzer} mid-infrared with submillimetre
flux densities, each being sensitive to a different temperature regime
(stars, hot dust, cold dust), can do a great deal to reveal the likely
power sources of the infrared-bright galaxies within the blobs. In
particular, a simple color-color plot (ratio of MIPS 24\m\ to 850\m\ sub-mm versus 
ratio of MIPS 24\m\ to IRAC 8\m ) can be a powerful tool for
discriminating AGN from star formation, as well as getting a picture
of the specific star formation (SFR per unit mass). 
   
In Figure 2 we plot the ratio of 24\m\ to 850\m\ sub-mm versus ratio of 24\m\ to
8\m\ for the three fields listed above that have both blobs and deep sub-mm
coverage. Any source without a 24\m\ detection is not plotted. In the
case of LAB6\_J2143-4423 there are two MIPS 24\m\ 
sources with confirmed $z$=2.38 redshifts \citep{sca09}), so
both sources (LAB6a and LAB6b) are included on the plot, but the submm
detection has been assigned to LAB6b\_J2143-4423 (as described above). Similarly,
LAB1\_J2217+0017 is associated with two equally bright 8\m\ sources (LAB1a and 
LAB1b). Due to the large sub-mm beamsize it is not possible to deduce
which source is likely associated with the 850\m\ flux and, of course,
both may be. We have therefore combined the 8\m\ and 24\m\ fluxes of 
LAB1a\_J2217+0017 and LAB1b\_J2217+0017 into a single source for the purpose of this
plot. If all the sub-mm flux were to originate in the slightly fainter
24\m\ source (LAB1a\_J2217+0017) the 24/850 \m\ ratio would go down by an
additional factor of 3, with very little change in 24/8\m . 

We overplot the same CE01/Mrk231 mixed models as used in Figure 1. We 
marked the ratio of star formation luminosity to AGN luminosity for
one of the models, running from 1.0 (100\% starburst) upwards to 0.0
(100\% AGN) contribution from star formation to the total bolometric luminosity.

All eight of the sub-mm detected components for LABs
have mid-IR/sub-mm ratios consistent with star formation.  An
additional six of the \Lya blob components are only upper limits,
but only two of those (LAB6a\_J2143-4423 and LAB19\_J1714+5015) are clearly outside and above 
the locus of star formation, consistent with IRS spectrum of LAB6a\_J2143-4423
\citep[see below,][]{sca09}  and the broad line nature of LAB19\_J1714+5015
\citep{pas96,col06a}. This submillimetre evidence for significant quantities of
cool dust in LABs was previously noted by \cite{web09}, who 
found that all but one of the SSA22 8\m\ sources they studied were detected at
850\m\ as well. That strong trend does not hold true for our sample,
where 6 of 14 of our LAB counterparts are detected at 8\m\ but not at 850\m\, but
that could be at least partially attributable to the slightly different depths
and redshifts in our study.  

This plot strongly indicates that the majority of these
infrared-bright components of LABs are not powered by AGN alone, but
there are two important caveats. First, the present sensitivity limits
of submillimeter surveys are just barely deep enough to conduct this
experiment. Not only are nearly half the sources not detected at all, but
none of the actual detections are that strong, ranging from 2.5-8
$\sigma$, with half of 4$\sigma$ significance or less. Fortunately the
expected range of  24/850\m\ ratio is large, especially at z=3.1,
allowing for a strong discrimination even with very large sub-mm
errors.

Secondly, while the correction for mass evolution is
likely generally correct, it is almost certain that there will be
real variations in specific star formation history from galaxy to
galaxy. On the plot this will move galaxies back and forth
horizontally, which will change what models and AGN/star formation
percentage one would associate with it. This is not an
important effect at $z$=3.1, where the models are less degenerative and there is
large separation in 24/850\m\ ratio as the AGN contribution is increased,
but it can be a concern at $z$=2.4. The mid-IR to
sub-mm ratios at the $z$=2.4 are not as strong for
discriminating the energy source because of the presence of the powerful
PAHs in the 24\m\ filter, which can somewhat mimic the rising power
law slope of an AGN. 

For instance, LAB18\_J1714+5015 (labeled in Figure 2) appears to
lie roughly in the neighborhood of the 93-98\% star formation models.
However, a change in the specific star formation of the nearby models
by 30\% would shift it towards the AGN models such that it would 
only take a factor of \ab 2-3 change in the 24/850\m\ ratio (either the model or
through a measurement error) to make this source AGN-dominated. In
fact, its IRS spectrum (see below) suggests that it is completely
AGN-dominated. To similarly shift $z$=3.1 sources would require
changes of factors of 10-20 in the 24/850\m\ ratio.
   
For further comparison we also plot sub-mm galaxies spectroscopically confirmed at similar
(\p 5\% in 1+$z$) redshifts from \cite{pop06}. The mid-infrared colors and
mid-infrared to sub-mm ratios for the \cite{pop06} sub-mm
galaxies are very similar to those from the LAB sample, with the
exception of one $z$\ab 2.4 sub-mm source (GN22) with a 24/850\m\ color so
cool it can not be easily explained by our simple set of
models. Three of these $z$\ab 2.4 sub-mm galaxies also have IRS mid-infrared
spectroscopy \citep[GN04, GN05, and GN19;][]{pop08}. All three have significant PAH
features, despite the fact that two of the three are detected in the
hard X-rays. Looking at the line-to-continuum ratios, \cite{pop08}
found that only GN04 (the source with the largest 24/850\m\
ratio included from their study) had $>$50\% contribution to the mid-infrared
energy output and they classified it as a mixed (AGN + starburst)
source. 

We note that one blob in our sample falls on the low end of the models. This is
the $z$=3.09 \cite{ste00} LAB1\_J2217+0017, which has a large
reported sub-mm flux \citep[16.8\p 2.9;][]{cha04}, but is quite
faint at 24\m\ with a flux of only 70\p 10 $\mu$Jy. This is actually
the combined 24\m\ flux of the two bright 8\m\ sources identified at
this location \citep{web09,gea07}. If split back into their two components
the 24\m\ fluxes are 25 and 47 $\mu$Jy, so if the submillimeter flux came from just one
of them the 24/850\m\ ratio would be even more extreme. LAB1\_J2217+0017 was
undetected in high spatial resolution observations (\ab 2$\arcsec$)
taken with the Submillimeter Array \citep{mat07}, indicating
that the 850\m\ flux must be widely spread out or split into multiple
sources. Alternatively, the original measurement might be a several
sigma deviation. ASTE-AzTEC 1.1 mm observations failed to detect it 
down to 10 mJy \citep{koh08}. However, even a factor of 10 less
flux would still place LAB1\_J2217+0017 in the region dominated by star formation. 

\subsection{PAH Features in LAB ULIRGs}

Of the six sources associated with LABs targeted with IRS
spectroscopy, four show significant PAH features. In order to
measure individual PAH emission lines, we used the IDL PAHFIT software 
package \citep{smi07a}, which fits all PAH emission lines and
continuum simultaneously. Given the limited coverage of 
the data, we only include the PAH lines in the fit that cover the
wavelength range of interest (6-9\m ) and a single red 
continuum. We chose this over the more direct method of isolating the
region directly around each line of interest and fitting a line and
continuum there \citep[i.e.,][]{pop08}, as that can underestimate the
equivalent widths by factors of up to 4. Fits to all six spectra are
shown in Figure 3 and all fluxes, equivalent widths, and
line-to-continuum ratios are listed in Table 3. 

Two of the spectra show no apparent PAH features, with all line fits
producing fluxes below the 1-$\sigma$ uncertainty. These
featureless sources are the counterpart LAB6a\_J2143-4423
\citep[previously reported in][]{sca09} and LAB18\_J1714+5015 \citep{kee99}.

The remaining four sources all show the 7-9\m\ PAH complex (\ab
7.4-8.7\m ) at the expected location for their
redshifts. The weakest PAH detection is that of source LAB1\_J2143-4423, with a 7.7\m\
feature (defined as the combination of the 7.60 and 7.85\m\ PAH
lines) detection of only 2$\sigma$ significance. However,
the whole 7-9\m\ complex for the LAB1\_J2143-4423 source is detected at 4$\sigma$ 
significance. Combined with their discovery at the appropriate
wavelengths there is little doubt the PAH features for LAB1\_J2143-4423 are
real.      

We measured a 7.7\m\ line-to-continuum (L/C) ratio for each source in order
to test the contribution of any possible AGN to the total infrared
energy output. For ease of comparison we use the same method of
\cite{rig99}, taking the average intensity of the line and
continuum over 7.57-7.94\m . The error in line intensity is just a
quadratic sum of uncertainty in the data, but the error in continuum
is dominated by the quality of the continuum fit, which has only two
parameters, temperature and a normalization factor, for each of which
the fitting routine returns an uncertainty. Using a Monte Carlo method
we take the two derived probability distributions of temperature and
normalization factors to create a large distribution of average continuum
intensities from which we measure a final rms. This continuum error is
consistently large, often dominating the error in L/C.   

Using their L/C definition, \cite{rig99} found that
AGN have L/C$\simeq$ 0.2, pure starbursts have L/C$\simeq$ 3, and the typical
local ULIRGs have L/C$\simeq$ 2, indicating they are almost completely star
formation dominated. The two sources from our suvey with no detected PAH features
have L/C ratios $<$ 0.2-0.3 (2$\sigma$ limits), clearly AGN-dominated sources.       
The two with the most powerful PAHs have L/C=5-6. Even accounting for large
errors in L/C (the LAB7\_J2143-4423 error is \ab 50\%), it is clear these are star
formation-dominated sources. The final two sources are a bit less
clear. With a L/C of 0.8\p 0.2, LAB1\_J1434+3317 appears to have its main mid-IR
contribution from an AGN, but the star formation component is far from
negligible. For LAB1\_J2143-4423, its L/C (1.5\p 0.7), suggests a similar but
opposite situation: the primary contribution is from star formation,
but the AGN component is significant. The large error on the LAB1\_J2143-4423 L/C
does add some further confusion, as a one sigma deviation could easily
change the likely power source. 

Another similar way to approach this problem is to look at the PAH
equivalent widths (EW). We plot the 6.2\m\ EW vs. the 7.7\m\ EW in
Figure 4. The 6.2\m\ feature is less contaminated by nearby PAHS and
silicate absorption and therefore can be less vulnerable to the
continuum model chosen. Unfortunately, for most of our data the 6.2\m\ feature
lies near the noisier wavelength edge of the LL1 IRS detector,
limiting the information available for the continuum there. The 6.2\m\ PAH
was also not always strongly detected for similar reasons.
In the case of LAB1\_J2143-4423, the 6.2\m\ PAH emission is undetected, so
instead we plot a  2$\sigma$ upper limit. 

Equivalent widths of PAHs are extremely vulnerable to how the exact
line fitting is done, as they are a combination of both the line
measured and continuum fit assumed. A factor of 2 difference in
measured fluxes can easily result in a factor of 4 difference in equivalent
width. This can make it problematic to compare equivalent widths
between studies that have not approached the measurement in a similar
way. We therefore compare our four PAH-detected sources to the $z$=1-3 ULIRGs of
\cite{saj07}, where they also applied a multi-component
PAH profile fit. Some differences remain, as \cite{saj07}
covered a longer range of wavelengths, which produces somewhat
different continuum fits. Our fits seem to be
favoring a continuum that is slightly lower, leading to slightly
larger EWs.  However, considering the large errors, the two methods
produce similar results.

The dashed lines on the figure (6.2\m\ EW=0.2\m\ and 7.7\m\ EW=0.8\m ) are
also from \cite{saj07} and represent suggested dividing lines
between strong-PAH (i.e. star forming) and weak-PAH sources
(i.e. AGN-like). Only 25\% of the \cite{saj07} high-z ULIRG sample fell
into the strong-PAH (they are half the sources on the plot, but that is misleading as
7.7\m\ limit-only objects are not plotted), as opposed to half of our
sample (our two PAH-free sources are not plotted). While the
lines are mainly a device to assist comparison between studies,
they do appear to be in rough agreement with what we found looking at 
7.7\m\ L/C ratios. 

Star formation rates can be derived from PAH luminosities
\citep[i.e.,][]{cha01,bra06},  although the uncertainties and unknowns increase with redshift and
luminosity. Using the 7.7\m\ to total infrared luminosity formulas
suggested by \cite{pop08} for their sample of submillimeter galaxies, we derive star formation rates
and/or limits for our six sources (see Table 3). As already discussed
above, our method of fitting all PAH features and continuum
simultaneously produces larger PAH fluxes than if one just fit a
continuum locally around each PAH emission feature, as \cite{pop08} did for their sample. Therefore if we simply put our
derived luminosities into the \cite{pop08} formulas we would
certainly be overestimating the star formation rate. For our 4 detected
7.7\m\ emission features, we found a mean correction factor to our fluxes of
1.7 if we use a method similar to that of \cite{pop08}. We
apply this correction, dividing the 7.7\m\ luminosities in
Table 3 by a factor of 1.7, before applying the formulas of \cite{pop08}. 

We find star formation rates in our PAH-detected sample ranging from
420 to 2200 M$_{\odot }$ yr$^{-1}$, significantly larger than the limits
we found for our two non-PAH sources which were $<$ 130-140 M$_{\odot
}$ yr$^{-1}$. While these rates are derived from
the 7.7\m\ fluxes, the 6.2\m\ feature can also be used to
derive total infrared luminosity and, consequently, SFR. While our
6.2\m\ feature detections are generally quite weak, the star formation
rates derived from them agree with those from 7.7\m\ within the
errors. The most discrepant source is that of LAB1\_J1434+3317, which despite
having a low L/C ratio does have the brightest 7.7\m\ luminosity of the
sample. While technically within the errors, its 6.2\m\ luminosity
points toward a SFR closer to a 1000 M$_{\odot }$ yr$^{-1}$. None of
the other 6.2\m\ SFR estimates are off by more than a few hundred M$_{\odot }$
yr$^{-1}$. This could be due to LAB1\_J1434+3317 being such a strongly mixed (starburst and AGN)
source or perhaps because it is such a highly luminous mid-infrared
source. Both are known to affect conversion from PAH luminosity to far infrared luminosity
\citep{mur09}. 

We can then ask the question whether these star formation rates are
enough to power the LABs. Star formation might cause the LABs in two
different ways: photoionization from Lyman continuum photons escaping
from the galaxy and supernova-driven outflows into the surrounding
medium, producing shocks.

For photoionization we can estimate the number of ionizing photons
produced from the \cite{bru03} models which use a high mass cutoff of
100 M$_{\odot }$. Assuming an age greater than 15 Myr (after which the
number of ionizing photons produced by continuous star formation
become roughly constant), for every 1 M$_{\odot }$ yr$^{-1}$ the
young stars will generate \ab 9$\times$10$^{52}$ ionizing photons
s$^{-1}$.  The predicted \Lya luminosity is therefore:

\begin{displaymath}
L_{Ly\alpha }=0.6f_{esc} h\nu \times [SFR] \times (9\times 10^{52}) \mbox{ ergs s}^{-1}
\end{displaymath}

\noindent
where h$\nu$ for \Lya is 2.58$\times$10$^{-12}$ ergs, 0.6 is the
fraction of absorbed Lyman continuum photons that will be re-emitted as \Lya
photons, and f$_{esc}$ is the fraction of Lyman continuum photons that
can escape from the galaxy. For a SFR of 1000 M$_{\odot }$ yr$^{-1}$
we predict 1.4$\times$10$^{44}f_{esc}$ ergs  s$^{-1}$.  For
$f_{esc}$ near unity one could power most blobs with the ionizing
continuum from the starbursts, but such a high escape fraction for the
Lyman continuum is highly improbable for these sources where most
of the energy is coming out as reprocessed dust emission. The majority of rest
frame ultraviolet photons are certainly not escaping along our line of
sight. If we apply a more conservative $f_{esc}$ = 0.1 then there are
not enough ionizing photons from star formation escaping to produce
any of the LABs with PAH-emitting sources presented here, with
deficits ranging from factors of 2 to 13. 

For supernova-driven outflows we start by estimating the number of
supernovae generated per year. Again using the \cite{bru03} models we
estimate 5$\times$10$^{-3}$ SN per M$_{\odot }$ yr$^{-1}$. If each
supernova produces \ab 10$^{51}$ ergs, then star formation is
generating roughly 1.7$\times$10$^{41}$ ergs s$^{-1}$ for every
M$_{\odot }$ yr$^{-1}$. This is enough energy to power the LABs, but
we still have to account for the efficiency of the conversion of the
supernova energy into kinetic outflows. \cite{tho98} indicates that as
much as 70--90\% of the supernova energy will be radiated away, leaving
only 1--3$\times$10$^{50}$ ergs of kinetic energy remaining. If we
account for this efficiency our typical 1000 M$_{\odot }$ yr$^{-1}$
infrared LAB counterpart will generate 2--5$\times$10$^{43}$ ergs
s$^{-1}$. Except for the most pessimistic assumptions, there appears
to be enough supernova energy available to power two of the three
star formation dominated LABs, with only LAB1\_J2143-4423 falling short by
nearly a factor of 4.  We note that a more top heavy IMF with larger
mass upper limits will produce both more supernovae and ionizing
photons. While this will certainly produce more \Lya photons,
it is unlikely the IMF could be radically different enough from our
assumptions to change the results. So while star formation can not
quite photoionize enough photons to generate LABs, there is enough
energy in supernova outflows to do so, but not in all cases.  

The requirement for detection at 24\m\ clearly has the potential to introduce a
bias into our analysis. Without a powerful 7.7\m\ feature the 24\m\ flux of $z$=2.4 sources
would generally be less, likely removing it from this sample ($>$0.2
mJy) altogether. One known $z$\ab 2.4 LAB, LAB19\_J1714+5015
\citep{kee99}, was excluded because its 24\m\ flux density is
not above \ab 0.2 mJy, just missing our cut. It is a known broad line AGN
\citep{pas96,col06a} which likely dominates its mid-infrared
output. However, that is the only 24\m -observed LAB at redshift $z<$3
known at the time of our program that was not
observed by IRS. Unfortunately the number of known LABs at redshifts
below $z$=3 remain very small (roughly a dozen), so we must
not draw too many conclusions from such a small
sample. Of those galaxies associated with LABs that are bright at
24\m\ , half have strong PAHs and appear to be star formation dominated.

None of the LABs in the higher redshift $z$=3.09 SSA22 field had
bright 24\m\ fluxes, but at their redshift the powerful 7.7\m\ PAH
feature moves beyond the MIPS 24\m filter which instead becomes sensitive to the
continuum below 6\m . That is a region of the spectrum of galaxies
that is generally dominated by hot dust, meaning at this higher
redshift one might expect a reverse of the selection bias: brighter 24\m\
will indicate a stronger hot dust component, i.e. a greater contribution
from AGN. Four of the five SSA22 blobs (80\%) with luminous X-ray
counterparts \citep{gea09} are detected at 24\m\ (40-160 $\mu$Jy), as
opposed to the remaining ten non X-ray luminous sources observed
with MIPS 24\m , for which only four (40\%) are detected (50-80$\mu$Jy).
 
We wish to provide one final word of caution on the use of PAH flux to
continuum ratios for trying to breakdown the total bolometric output
for such powerful, high-redshift objects. It is entirely possible that star forming galaxies,
forming at rates of more than 1000 M$_{\odot}$/yr might produce significant quantities of hot
dust emission without AGN \citep[i.e.,][]{hun02} and/or produce conditions that destroy
the grains that produce PAHs \citep[i.e,][]{gal03}. However, the continued
discovery of PAHs in bright, high-z sources \citep[i.e.,][]{fad10,hua09,yan07} suggests
the issue of PAH destruction is likely not a serious problem. If anything, there appears to be
evidence that the brightest, high-redshift galaxies are over-producing
PAHs compared to their lower redshift, less luminous counterparts
\citep{mur09}.      

Whether we measure {\it Spitzer} photometry ratios, compare mid-infrared to
submillimeter fluxes, or examine the strength of PAHs using IRS spectroscopy, we
consistently find a very similar answer: half to 2/3 of all LAB
counterparts appear to have their infrared luminosity powered mainly by star formation. This does
not prevent powerful AGN from also existing within these sources or rule AGN out as a
source of the extended \Lya emission. However, with such large
reservoirs of starburst energy it does make star formation-driven
LAB models, such as superwind outflows, a much more likely power
source for the majority of LABs. 

\subsection{Blob Counterpart Masses}

Several lines of evidence suggest that LABs mark
regions of massive galaxy assembly: their location at the peak
of high redshift structures \citep{mat09,mat04,pal04}, their number
densities comparable to galaxy clusters in the nearby and
high-z universe \citep[10$^{-5}$--10$^{-6}$ Mpc$^{-3}$;][]{yan09},
the freqent indicators of merger and/or interaction \citep{col06b,cha04,fra01}. One
possiblity is that they could be a phase in the formation process of
massive elliptical galaxies themselves. {\it Spitzer} IRAC data
provides us with rest wavelength measurements of the 1-2\m\ portion of
the spectral energy distribution (SED) of the infrared
components of the LABs being analyzed in this paper. This is ideal for
estimates of mass, lying at the peak of stellar light output while
minimizing the effects of dust extinction.

However, the mass-to-light ratio (M/L) can change significantly, even
at these wavelengths, for the young objects that one might expect to find
in the early universe. Robust mass estimates require a good
age estimate, so therefore most studies perform a full SED fit to
estimate the age in order to derive a M/L ratio and total mass.  

Previous measurements of LAB counterpart masses regularly find
the brighter LABs to be around 10$^{11}$ M$_{\odot }$
\citep{gea07,smi08}. \cite{uch08} measured masses for seven SSA22 LABs,
deriving masses from their K-band data ranging from 4$\times$10$^9$ --
1.1$\times$10$^{11}$ M$_{\odot }$, which they found to be roughly correlated with
the luminosity of the LABs within which they resided. In addition, both
LAB1\_J2143-4423 \citep{fra01} and LAB6b\_J2143-4423 \citep{sca09} 
have undergone a full SED fitting analysis, producing masses of \ab
1.5$\times$ 10$^{11}$ M$_{\odot }$ and 4$\times$10$^{11}$ M$_{\odot }$
respectively. In the case of LAB1\_J2143-4423 there were two fits
done, one to each major component, but as we are unable to resolve the two
components in the coarser IRAC data used in this study we
report only the combined mass here.

A full SED fitting is not possible for all galaxies
associated with the LABs, due to the paucity of the some of the
necessary deep data -- especially the near-infrared -- for the different fields. For instance, the
4000\AA\ break, one of the more critical features for
age estimation, lies entirely in the near-infrared at the redshifts of
the LABs. More importantly, even with full spectral coverage, SED fits can suffer from
AGN-contamination and age/dust extinction degeneracy, both of which
will strongly effect any mass determination. Even measurements of the
4000\AA\ break, which are robust for measuring the age of older populations, will fail to
provide accurate ages for young starbursts, like those likely present
at the high energies and early universe epoch we are studying. We therefore decided to
examine the mass upper limits for our LAB counterparts.

Almost all of the galaxies in our study are covered with the four IRAC
channels, covering the rest-frame
1.1\m\  both at $z=3.1$ and at $z=2.4$. We convert the 1.1\m\ luminosity to stellar masses assuming a simple
single stellar population model \citep{bru03} with no dust. In order
to compare galaxies at different redshifts, 
we have computed maximally old stellar population
models at the observation redshift, assuming $z=8$ formation redshift. The derived masses, therefore, provide an upper limit
to the real stellar masses, since they are computed under the
assumption that all the 1.1\m\ luminosity is due to old stars. To give
an idea how much this could overestimate the masses we can compare to
the SED model fit done for the X-ray detected LABs in SSA22 of \cite{gea09}. They
found that a relatively young SED model
with 100 Myr of continuous star formation history and an extinction of
A$_V$ =1.5 was the best fit for their X-ray emitting sources. 
If this same model applied to all the LAB counterparts, their masses would be lower by a factor of
5.7 at $z$=3 and 6.7 at $z$=2.4 than our plotted maximum mass limits. However, for
the cases of a couple of our brighter sources (LAB1\_J2143-4423 and
LAB6b\_J2143-4423) the upper limits are within a factor of \ab 2 of the
masses previously derived from a full SED fit \citep{sca09,fra01}. Of the SSA22 LAB counterparts
included in both our study and that of K-band study of \cite{uch08},  the
typical difference is a factor of \ab 2-2.5, although two sources have very large
disparities (LAB7\_J2217+0017 \& LAB16\_J2217+0017), for which the difference is closer to a
factor of 8-10. The 4.5/8\m\ ratio of LAB16\_J2217+0017 suggests it may contain an
AGN, so these may be examples of AGN contamination. LAB7\_J2217+0017 is too faint
to apply any of our infrared AGN/starburst diagnostic ratios.

An analysis using these mass upper limits has another advantage. If
the LAB components are all of similar formation era, whatever that may be, 
then they should have very similar 1.1\m\ M/L ratios and 
their masses {\it relative} to one another would be accurate no matter
when exactly they all formed. 

We plot these mass upper limits versus the \Lya luminosity of the blob
within which they are found in Figure 5. The errors plotted are just
from the photometry and do not represent the significant uncertainties
resulting from the model M/L ratios, choice of formation redshift, etc. In cases of more than one
mid-infrared source (LAB1\_J2217+0017 and LAB6\_J2143-4423) both are plotted with the same
\Lya luminosity. The clearest result is that there are no low mass,
high \Lya luminosity sources. In fact, one could split the blobs into
two groups: a \Lya bright ($>$2$\times$10$^{43}$ ergs sec$^{-1}$),
high mass sample and a \Lya faint ( $<$2$\times$10$^{43}$ ergs
sec$^{-1}$) sample with no strong preference for mass which does not
quite reach the most massive end of the LAB sample. This lack of
\Lya bright, low mass sources would continue to hold true even if one were to reduce
the masses of the bright \Lya sources by the factor of \ab 6
suggested for the \cite{gea09} X-ray LABS, 
while leaving the masses of the \Lya faint sources
unchanged. The limited mass fitting by ourselves and previous studies
\citep[i.e,][]{uch08} suggest that if there is any systematic
difference in the M/L ratio with \Lya flux, it is actually the \Lya
bright sources that tend to have higher M/L,  which would make 
this split stronger.  

This possible split is similar to
the two populations proposed by \cite{web09}, infrared
luminous and infrared faint, which they based mainly on whether the counterpart possessed
an 8\m\ detection. Their conclusion that the infrared
luminous LAB counterparts had large hot dust contributions from AGN and/or intense starburst ULIRG
activity while the infrared faint counterparts resembled cooler, pure star
formation systems, does not appear to apply as well to \Lya luminosity
or mass. Our plotted sample shows evidence for AGN contribution
throughout the entire range of \Lya luminosity studied and all but the
faintest end of the mass limits. However, the \cite{web09} study
does include the smallest and faintest of the SSA22 blobs which we
removed from our study. 

The presence of a bright AGN would contaminate the rest wavelength 1.1\m\ and produce
masses much higher than are actually present in stars. We have marked
all those sources with IRAC colors indicating AGN (this includes our
three AGN-dominated IRS spectra) in red. We also mark the two
borderline AGN cases from Figure 1 in yellow. Finally we circle all
the X-ray detected blob counterparts from \cite{gea09}. If these potential contaminant LABs were
removed from the sample, a weak trend
appears with the most massive sources associated with the brightest
\Lya luminosities. However, with only \ab 10 sources and the general
uncertainties in both AGN contamination and actual mass this trend is
far from confirmed. 

While it is unlikely that the masses of these mid-infrared components all lie at
their upper limits, it is clear many are quite large, around
10$^{11}$ M$_{\odot }$. Massive galaxies with substantial
($>$0.5 Gyr) ages are not unusual at these redshifts \citep{kri08}. Some of our largest masses could be a combination of
a close pair, like those known to be in LAB1\_J2143-4423 \citep{fra01}. In
another case, LAB1\_J2217+0017, the two sources are barely
distinguishable at the IRAC resolution, so it would not be that
surprising to find others that could not be. Whether a single galaxy
or some sort of merging/interacting pair (or more), the total mass of
all the galaxies is likely indicative of the size of the potential well in which
the galaxy is assembling. The weak correlation of mass and \Lya
luminosity, if real, would indicate a correlation between the size of
this potential well and the energy source that is powering the
blob. 
If the LAB is a cooling flow one would expect a direct correlation of
potential well to \Lya luminosity, but that does not rule out stellar wind or AGN
illumination models, as a greater potential well might be expected to
drive more gas inflow causing star formation or feeding a supermassive
black hole.    

\section{Summary and Conclusions}

Mid-infrared {\it Spitzer} ratios (rest frame near and mid-IR)
indicate that \ab 60\% of LAB counterparts are consistent with being
cool starbursts, while the rest have a substantial hot dust component
that one would expect from an AGN, although extreme
starbursts are a possibility in some cases. Including submillimeter observations (rest frame far
infrared) in the analysis comes to a similar conclusion: roughly 2/3
of LAB counterparts are consistent with the total bolometric energy
output being dominated by star formation. 

IRS spectroscopy of six of the brighter (and lower redshift) sources
found 4 of 6 to have measurable PAHs. The other two were featureless power law
spectra indicative of AGN-domination. Of the four detected, two had
L/C ratios and PAH EWs suggestive of mixed sources, with
energy contributions from both star formation and supermassive black hole
accretion. 

In general, the stellar masses of the LAB counterparts are quite large, around 10$^{11}$
M$_{\odot }$. There is a weak trend with the \Lya luminosity of the
host blob. This could be suggestive of two populations of LAB: one
\Lya luminous and generally massive and one fainter and slightly less massive,
but generally covering a wide range of stellar masses. Alternatively, the LAB
counterparts could be one continuous population, with mass growing with
\Lya luminosity. Indications of AGN are seen at all \Lya luminosites
and all but the smallest masses.   

A lot of the work on LABs has been spent trying to determine the
energy source that powers them. Not only would that allow us to
understand the physics of these giant clouds of extended \Lya
emission, but could possibly provide valuable information on the
assembly of massive galaxies, including questions of AGN feedback,
escaping ionizing radiation, and/or cooling flows. It has been
theorized that LABs could be a short lived evolutionary step in the life of most
galaxies at these redshifts \citep{gea09}. The \Lya halo could
be powered by star formation superwinds, growing larger and larger
until the central AGN grows enough to blow out most of the gas and 
cutting off the LAB's power source \citep{web09}.  

The problem with trying to place all the observed LABs in an
evolutionary sequence is that again and again the blobs resist
efforts to link them to a single power source. 
While \cite{gea09} found five LABs were strong X-ray sources,
their further submillimeter analysis suggested that even for these objects, the total bolometric output for
the galaxy is dominated by star formation. The presence of AGN,
seen in many LABs, could just be confusing the analysis or maybe
indicating a future evolutinary stage. Our own analysis of submillimter
and mid-infrared data suggests a great deal of the LAB infrared
counterparts show no indication of a significant hot dust component,
again pointing towards star formation, but there are several
significant exceptions. 

The IRS spectra from this study reveal two LAB infrared counterparts
each with an unambiguous, featureless AGN mid-infrared spectrum. Two
other sources have similarly unambiguous powerful PAH, star dominated
mid-infrared spectra. If the infrared counterpart is the power source
for the LAB, it is difficult to see how there could possibly be a
single explanation for what powers all \Lya blobs. Evolutionary
scenarios that leave the \Lya halo behind while the internal galaxy changes
over to a new energy source (like an AGN turning on) are probably not
viable due to the rapid cooling time of the ionized gas halo (\ab 1-2
Myr). We note that for LAB6\_J2143-4423 there is both an AGN and a star formation
dominated counterpart, so those particular IRS spectra do not
contradict the single power source model.

The estimated star formation rates for the PAH-emitting LABs generate enough
energy in supernova outflows to power two of the five LABs we
observed using IRS. Neither AGN nor cooling flows are needed to explain the \Lya
emission for these powerful PAH sources. More generally, cooling flow
models may deposit too much mass onto their galaxies, depending
on the exact duty-cycle of the \Lya halo \citep{gea09}. However,
there are several well studied, bright LABs with no obvious
infrared power source \citep{smi07b,nil06,pre09}, for which cooling
flows remain the most viable explanation. 

While more study is certainly required, we would suggest that the data
to date do not point to a single, uniform source of power for the \Lya
blob. Instead of being a homogenous group of objects, all created in
the same way, the LABs are likely a heterogenous group, with
different power sources depending on the object. What LABs likely all have in
common is their environment: the dense, gas-rich infall zones at the
centers of high redshift over-densities where the most massive
galaxies are being born. Outflows or photoionization from intense star
formation may drive the majority, but AGN almost certainly play a
important role in others. Cooling flows could account for only a small minority
of those seen, which would allow shorter duty-cycles and fewer
issues of mass deposition.

\begin{figure}
\includegraphics[angle=90,width=7in]{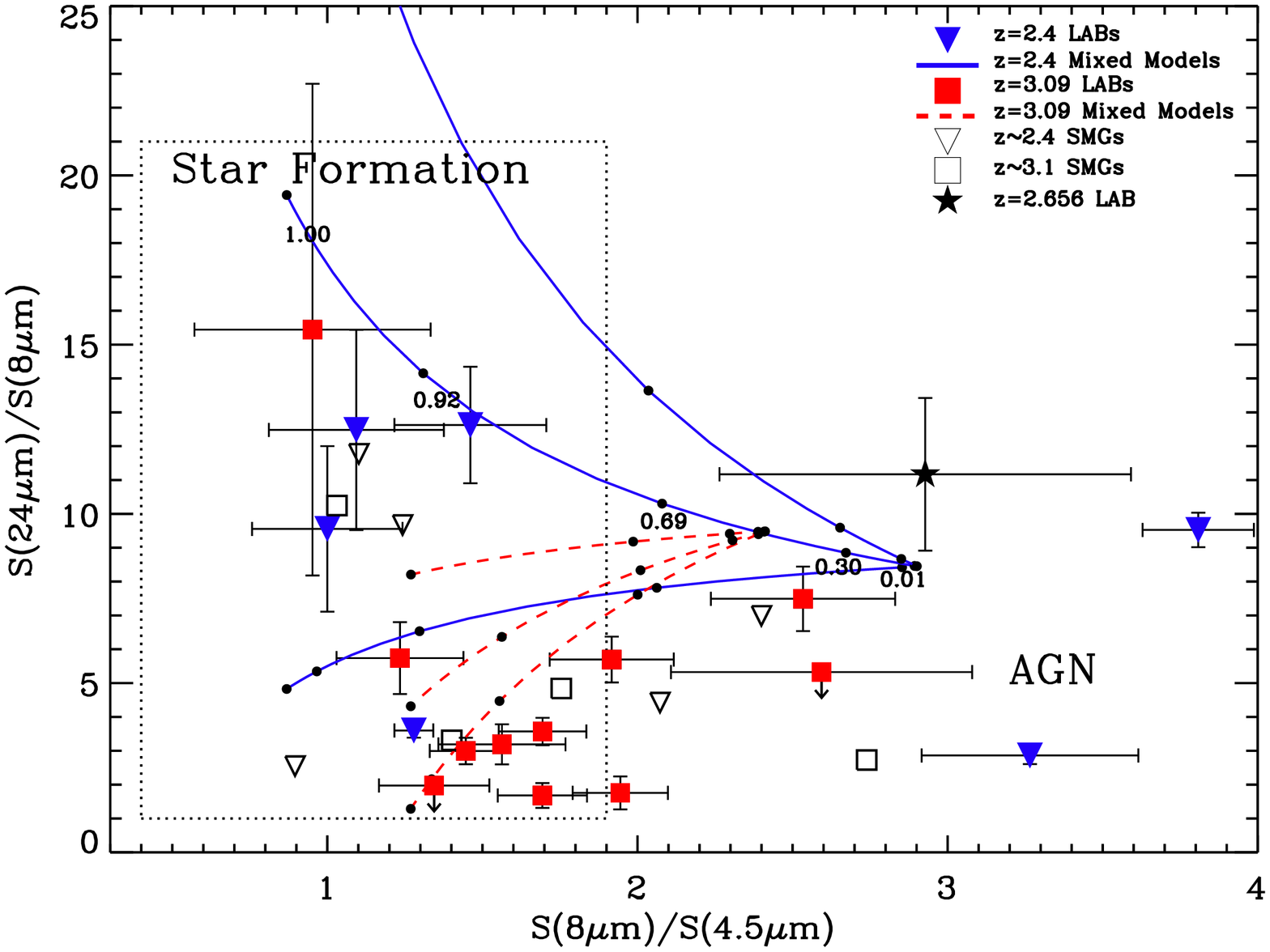}
\caption{Ratio of 8\m\ to 4.5\m\ versus ratio of 24\m\ to
  8\m\ for our sample of \Lya Blobs (solid symbols). Triangle (blue) symbols 
  are z=2.4 sources, squares (red) are z=3.1, and the star is
  LAB1\_J1434+3317 at z=2.66. For comparison we also
  plot the sub-mm galaxies (SMGs) at similar redshifts from \cite{pop06}
  as hollow symbols. The lines are models derived from \cite{cha01},
  running from star formation-dominated on the left to AGN 
  dominated at the right. The solid lines are z=2.4 (blue) models, while the
  dotted lines are z=3.1 (red). The rectangular box is taken from a similar
  plot of submillimeter galaxies from \cite{pop08}, marking the
  likely location of galaxies powered by star formation. } 
\end{figure}

\begin{figure}
\includegraphics[angle=90,width=7in]{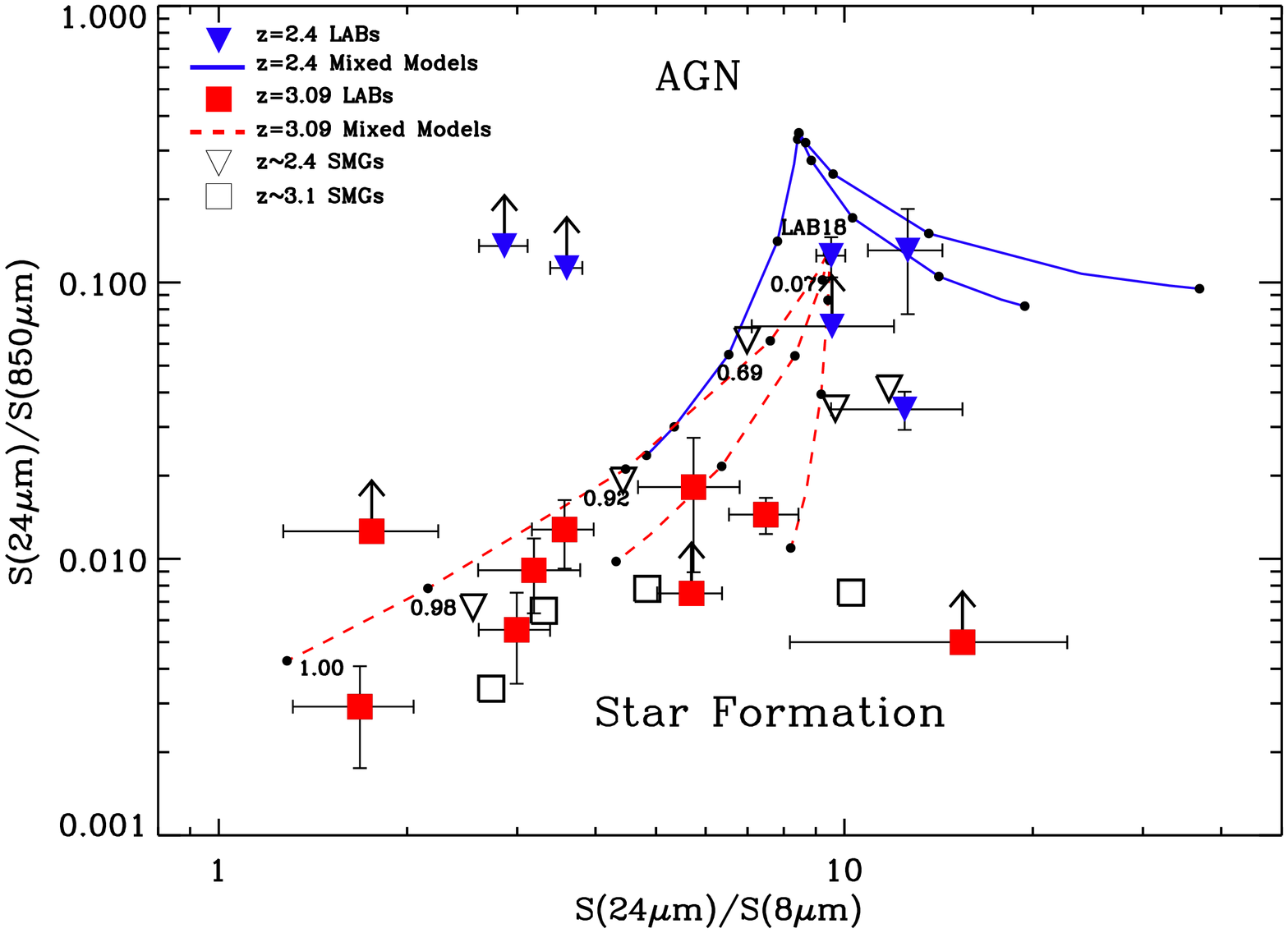}
\caption{Ratio of 24\m\ to 850\m\ sub-mm versus ratio of 24\m\ to
  8\m\ for our sample of \Lya Blobs (solid symbols). Triangle (blue) symbols 
  are z=2.4 sources, while squares (red) are z=3.1. For comparison we also
  plot the SMGs at similar redshifts from \cite{pop06}
  as hollow symbols. The model lines are the same as in figure 1, now running from
  star formation-dominated at the bottom up to AGN dominated at the top. All of the sub-mm
  detected \Lya blobs plotted appear to lie at the locus of star formation.} 
\end{figure}

\begin{figure}
\center{\plottwo{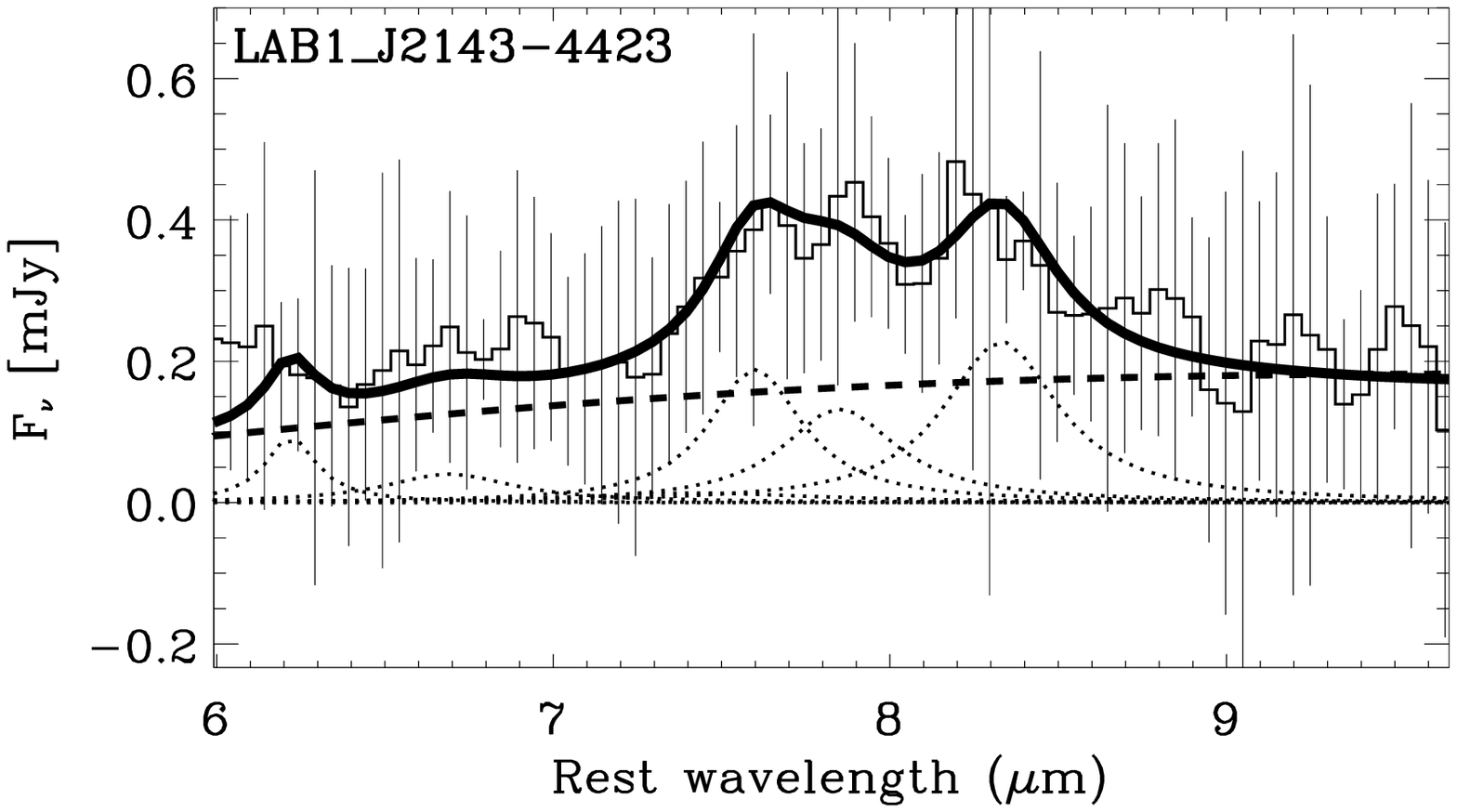}{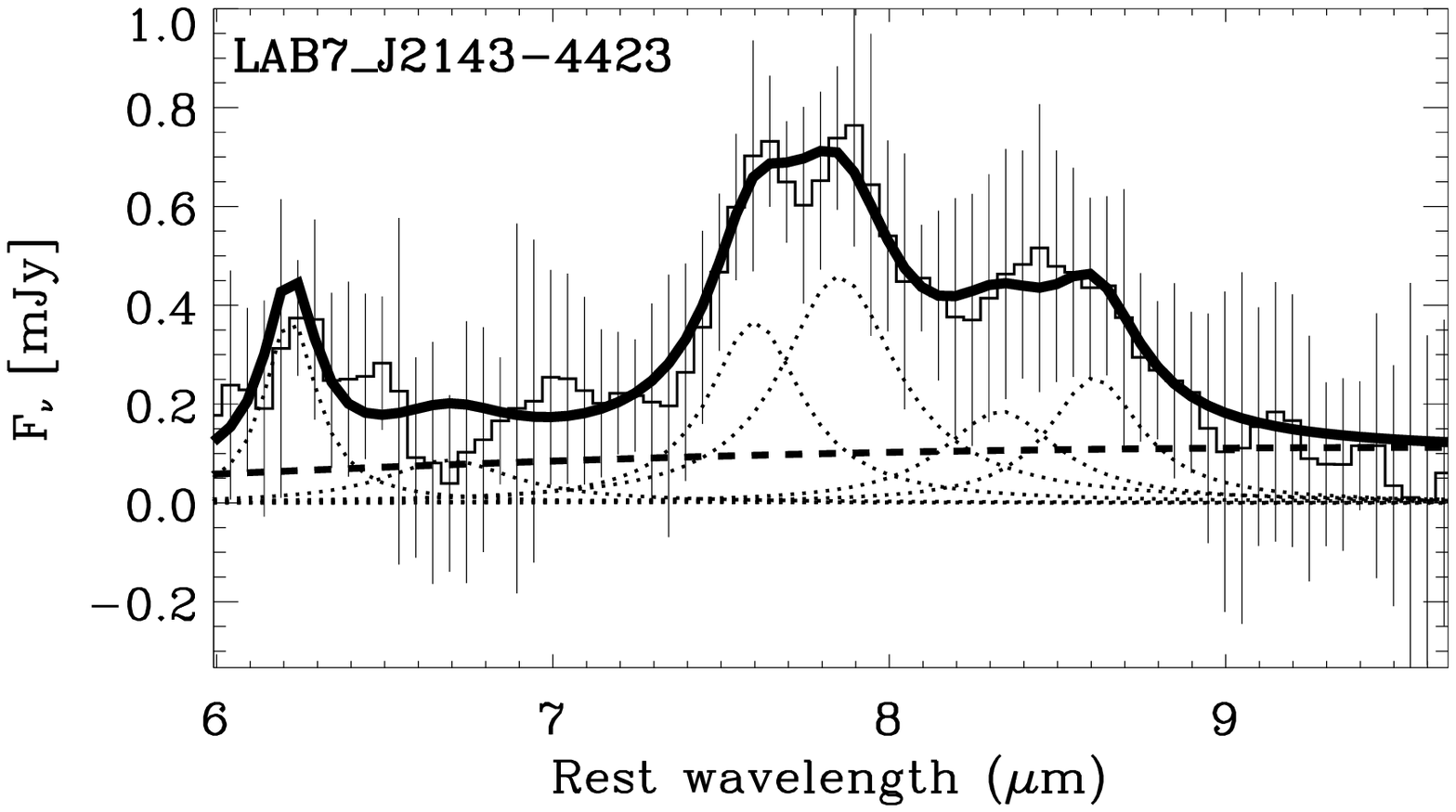}
\plottwo{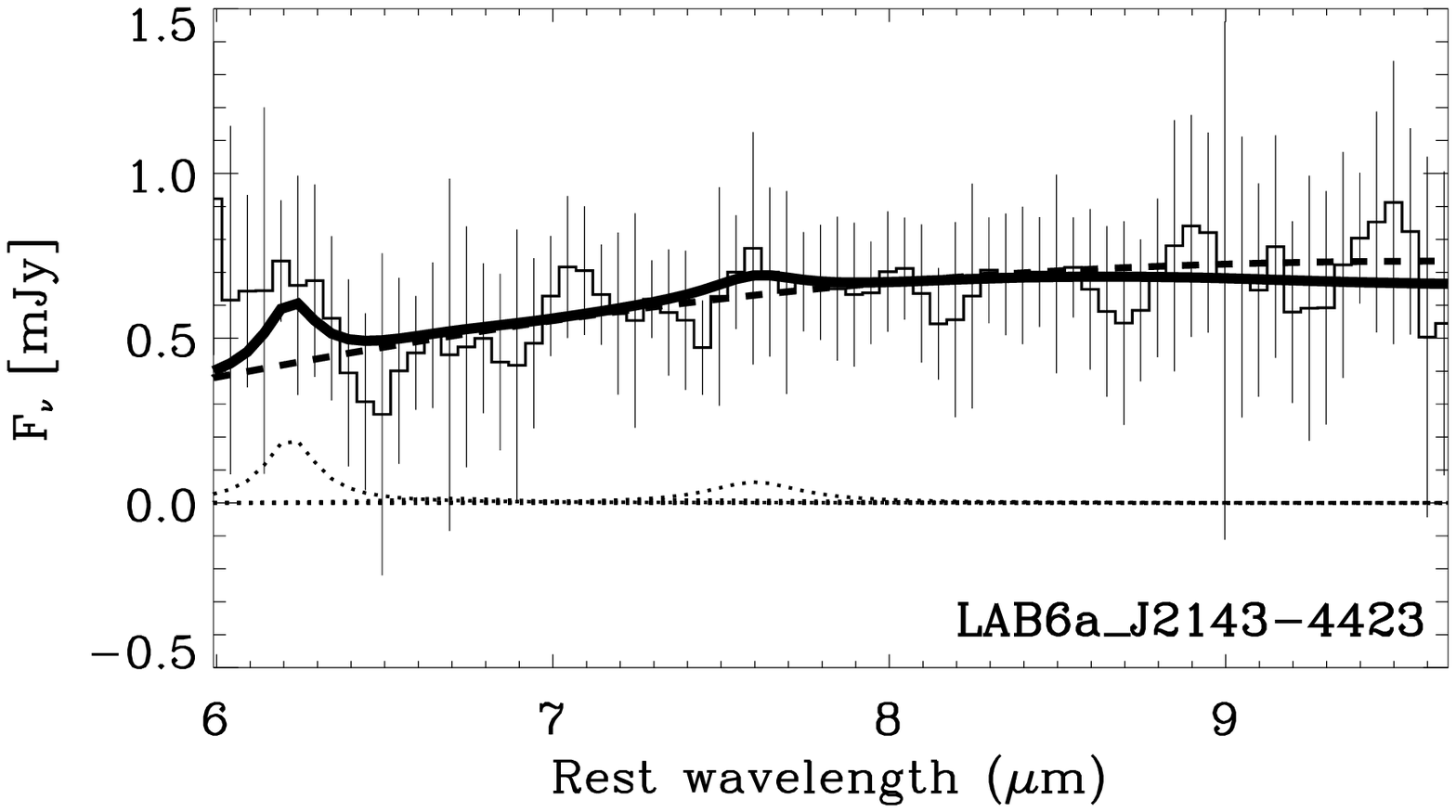}{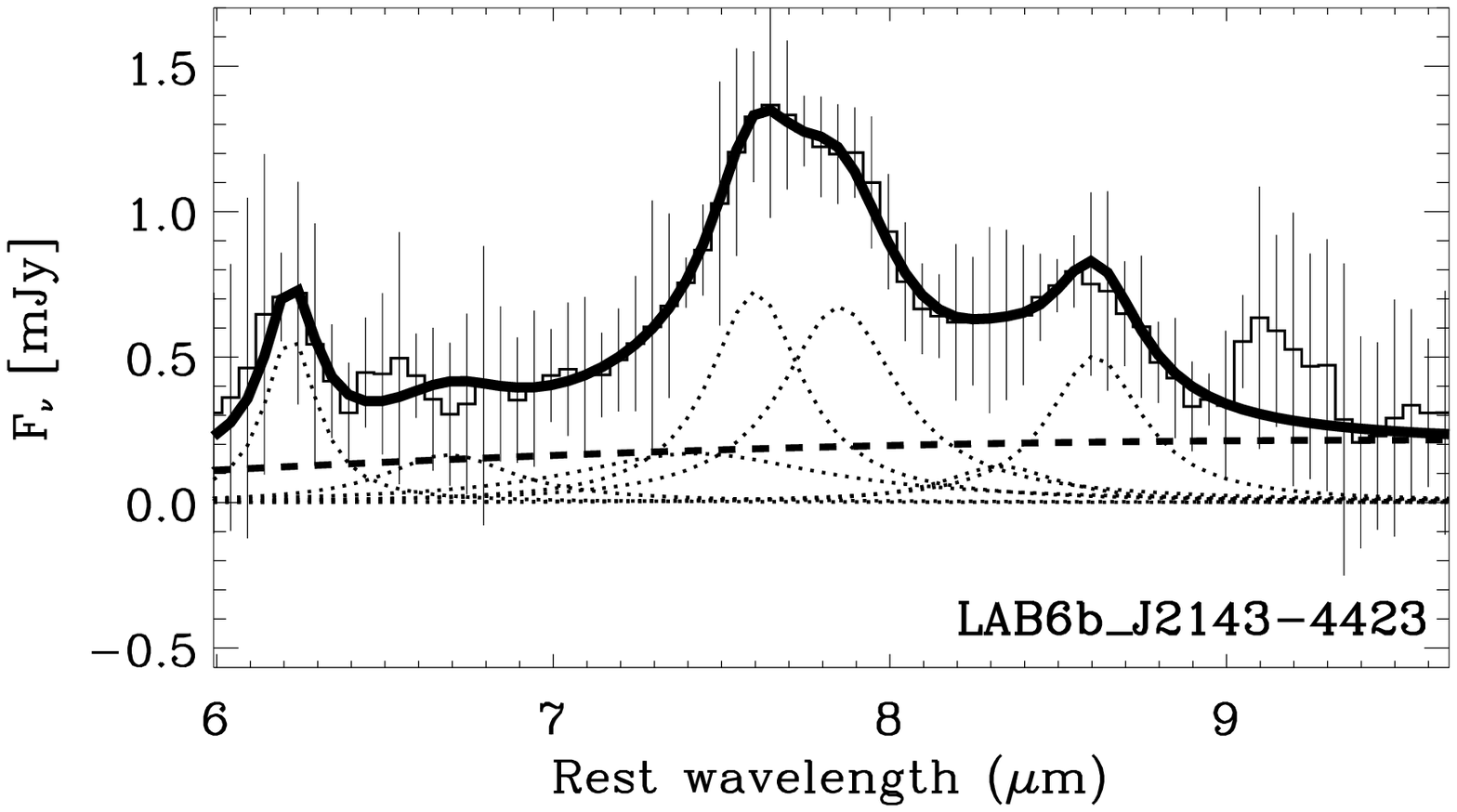}
\plottwo{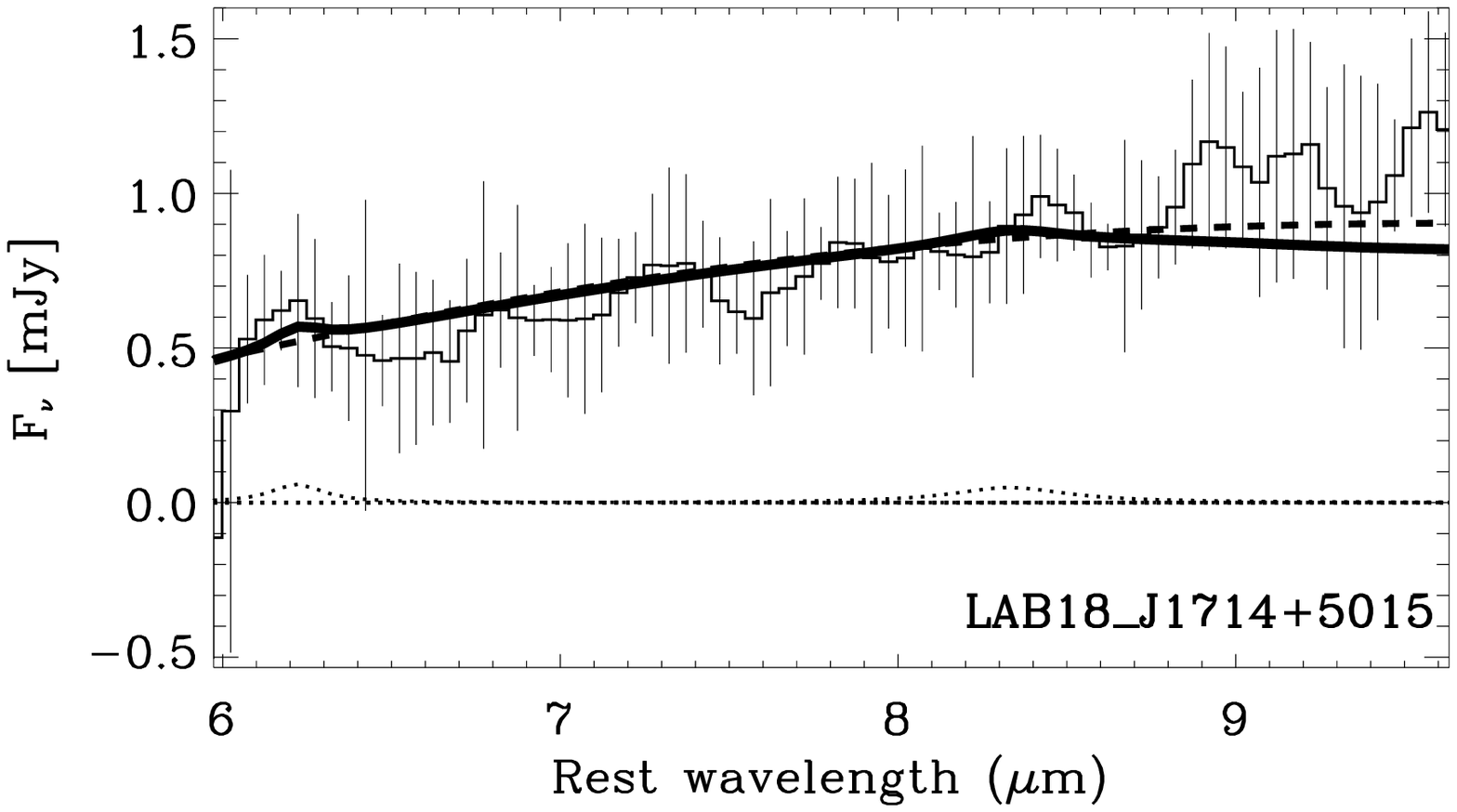}{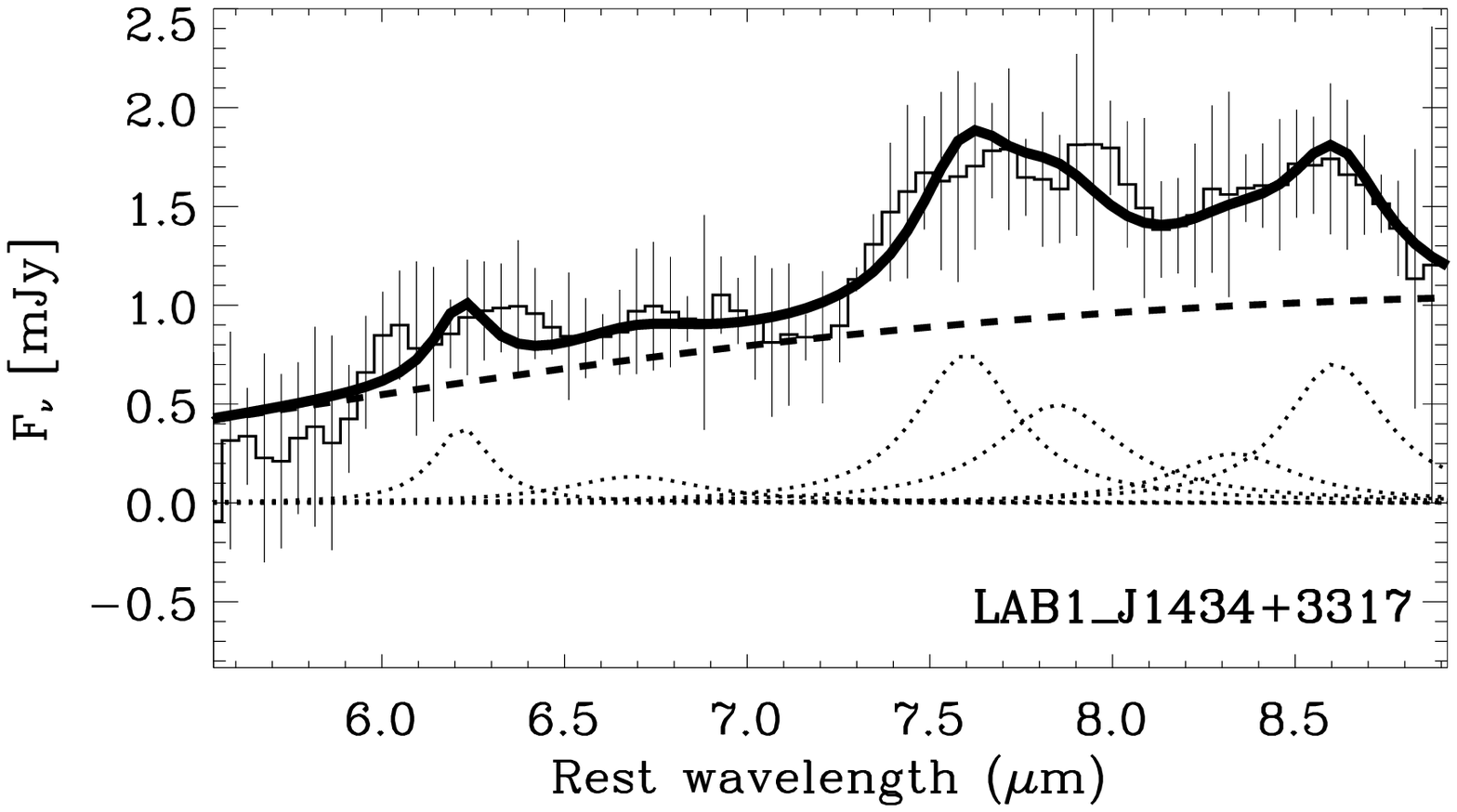}}
\caption{IRS spectra of six MIPS-detected LABs. Figures show best
  continuum (dashed line) and PAH fits (dotted lines) as determined by
  PAHFIT. The combined fits (PAHs + continuum) are the thick solid lines
  overlaid on the data.} 
\end{figure}

\begin{figure}
\plotone{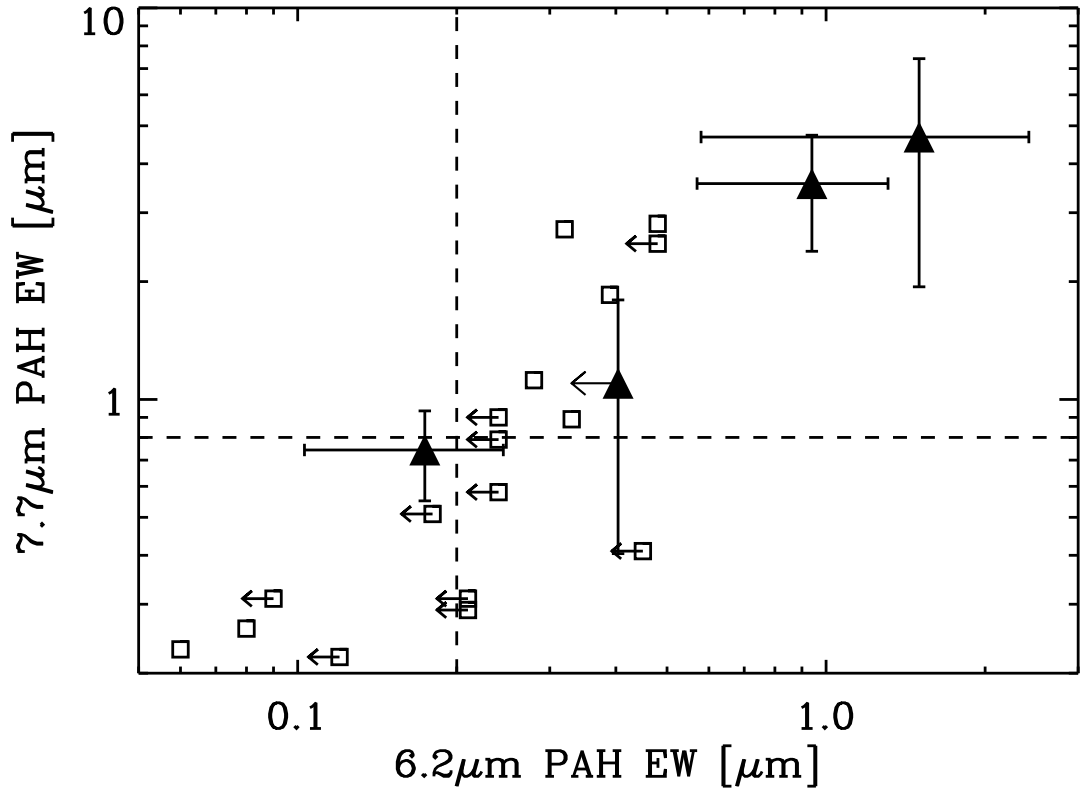}
\caption{6.2\m\ PAH equivalent width vs. 7.7\m\ PAH equivalent
  width. The square points are high-z ULIRGs from \cite{saj07}. The
  dotted lines at 6.2\m\ EW = 0.2\m\ and 7.7\m\ EW = 0.8 approximate a
  cut-off between AGN and star-formation dominated 
sources.}
\end{figure}

\begin{figure}
\includegraphics[angle=90,width=7in]{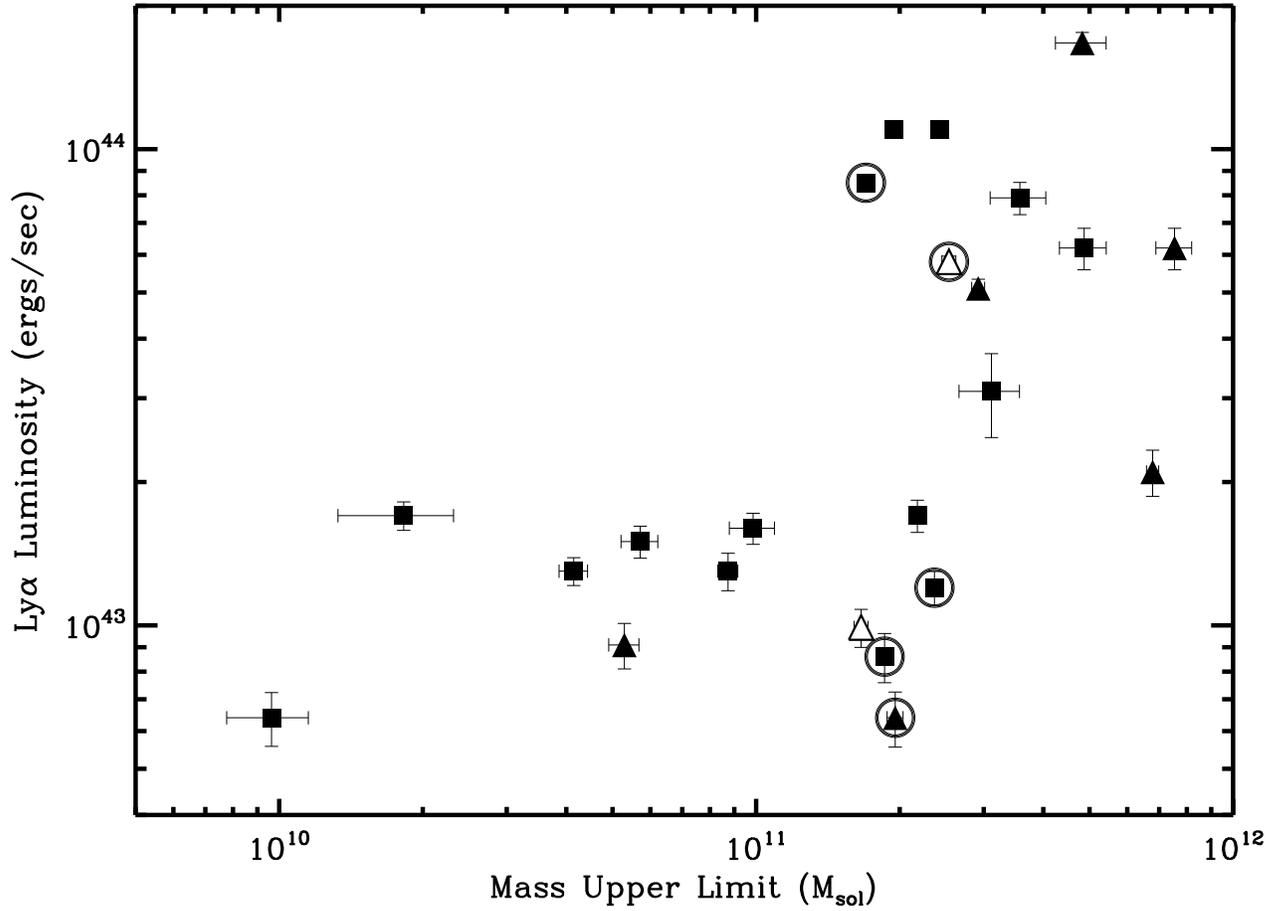}
\caption{Plot of mass upper limits versus \Lya luminosity for the
  associated blob. Blob components with AGN Spitzer colors
are plotted as solid triangles while the two with borderline 8/4.5\m\ colors are
plotted as empty triangles. The X-ray detected blobs are circled.}
\end{figure}

\begin{deluxetable}{lccccccl}
\setlength{\tabcolsep}{0.05in} 
\tabletypesize{\scriptsize}
\tablehead{
\colhead{Name\tablenotemark{a} } &
\colhead{4.5\m\ flux} &
\colhead{8\m\ flux\tablenotemark{b} } &
\colhead{24\m\ flux} &
\colhead{850\m\ flux} &
\colhead{\Lya Lum}&
\colhead{Mass Up. limit} &
\colhead{Previous Names} \\
& $\mu$Jy & $\mu$Jy & $\mu$Jy & mJy & 10$^{43}$ ergs s$^{-1}$ & 10$^{11}$ M$_{\odot }$ & }
\tablecaption{LAB Infrared Counterparts}
\startdata    
LAB1a\_J2217+0017 &  8.6  \p 0.3    &  14.6 \p 1.2    &  25 \p 5     & 16.8 \p 2.9 & 11 &2.0 & Blob 1, LAB01-a \\
LAB1b\_J2217+0017  & 10.7  \p  0.3   &  15.5 \p 1.2    &  47 \p 5 & 16.8  \p 2.9 & 11 &2.4 & Blob 1, LAB01-b \\
LAB2b\_J2217+0017\tablenotemark{c}    &  7.5 \p 0.2      &  10.1  \p 1.3  &  $<$10 &3.3 \p 1.2  & 8.5  & 1.7 & Blob 2, LAB02-b\\
LAB3\_J2217+0017    &  11.1  \p 0.4   &  21.5 \p 1.6   &  38 \p 10 & $<$3           & 5.8 &2.5 & LAB03\\
LAB5\_J2217+0017   &   9.5 \p    0.3   &  14.8 \p 1.9   &  47 \p 6 & 5.2 \p 1.4    & 1.7 & 2.2 &  LAB05\\
LAB6\_J2217+0017   &   4.3  \p   0.4   &  $<$9.4         &   n/a & $<$3.6       & 1.6 & 0.99 & LAB06 \\
LAB7\_J2217+0017  &    2.5   \p  0.2  &   $<$4.6         &   $<$24 &  $<$3.2      & 1.5 & 0.57 & LAB07 \\
LAB8\_J2217+0017   &  0.9 \p   0.1    &    $<$2.6         &  $<$10 & $<$10.6    &  1.7 & 0.18 &  LAB08\\
LAB9\_J2217+0017    &  3.6  \p  0.2   &   3.4 \p 1.4  &  53 \p 13   & $<$10.6   & 1.3  & 0.87 & LAB09\\
LAB10\_J2217+0017  &    n/a\tablenotemark{d}  &    n/a\tablenotemark{d}   &  207 \p 12 & 6.1 \p 1.4   & 2.2 & n/a &  LAB10\\
LAB11\_J2217+0017   &   2.3  \p  0.1  &  6.0 \p 1.1  &  $<$16 & $<$10.6      & 0.91 & 0.53 & LAB11\\
LAB12\_J2217+0017  &    8.2  \p   0.3 & 10.1 \p  1.6    &  58 \p 5     & 3.2 \p 1.6  & 0.86 & 1.9 & LAB12\\
LAB14\_J2217+0017  &    10.3  \p  0.3  & 17.5   \p  1.4   &  63 \p 5     & 4.9  \p 1.3  & 1.2 & 2.4 &  LAB14\\
LAB16\_J2217+0017   &   7.3 \p   0.2  & 14.0  \p  1.4   &  80 \p 5     & $<$10.6     & 0.99 & 1.7 &  LAB16\\
LAB18a\_J2217+0017\tablenotemark{e}   &   8.4  \p   0.3 & 21.2 \p   2.4   &  159 \p 10         & 11.0 \p 1.5 & 0.64 & 2.9 &  LAB18-a\\
LAB19\_J2217+0017   &   1.8 \p   0.1  &  $<$2.4          &  $<$24 & $<$10.6      & 1.3 & 0.41 &  LAB19\\
LAB20\_J2217+0017  &    0.5  \p  0.1 &   $<$2.4          &  $<$10            & $<$3.0       &  0.64 & 0.096 &  LAB20\\
LAB1\_J2143-4423    &  24.7 \p 3.3       & 24.7 \p 5.0      & 236 \p 37 & $<$1.7         & 7.9  & 3.6 &  B1 \\ 
LAB6a\_J2143-4423  & 58.0 \p 5.0        & 189 \p 12       & 542 \p 34         & $<$2.0       &  6.2 & 7.5 &  B6-Ly1 \\
LAB6b\_J2143-4423  & 34.7 \p 3.9        & 50.7 \p 6.3     & 640 \p 36 & 4.9 \p 2.0    & 6.2 & 4.9 & B6-Ly2 \\
LAB7\_J2143-4423    & 21.4 \p 3.1        & 23.4 \p 5.0     & 292 \p 30 & 8.4 \p 1.0    & 3.1  & 3.1 & B7 \\ 
LAB18\_J1714+5015 & 19.3 \p 0.6       & 73.5 \p 2.6     & 700 \p 28 & 5.6 \p 0.9     & 5.1  & 2.9 & Obj. 18 \\
LAB19\_J1714+5015 & 44.1 \p 1.3       & 56.4 \p 2.2     & 200 \p 9 & $<$0.9         & 2.1 & 6.8 &  Obj. 19 \\
LAB1\_J1434+3317   & 26.3 \p 3.1       & 77.0 \p 14.9   & 860 \p 50 & n/a               &  17 & 4.8 & SST24 J1434110\\
&&&&&&&    +331733
\enddata
\tablenotetext{a}{Primary LAB reference for each field -- J2217+0017 (SSA22):
  \cite{mat04}; J2143-4423: \cite{pal04}; J1714+5015 (53w002): \cite{kee99}; 
J1434+3317 (NDWFS): \cite{dey05}.}
\tablenotetext{b}{All limits listed are 2$\sigma$.}
\tablenotetext{c}{We found only one source with 8\m\
  flux bright enough to be a likely infrared counterpart. It had been
  previously labeled as counterpart "b'' \citep{gea07}. We keep the
  the "b'' label for this source, but there is no counterpart "a''
  used in our analysis.}
\tablenotetext{d}{The counterpart to LAB10 falls just off the IRAC channels 2 and 4
  fields, but not IRAC channel 1 (16.0 \p 0.5 $\mu$Jy) or channel 3
  (28.9 \p 2.0 $\mu$Jy).}
\tablenotetext{e}{We do not find the object "b'' \citep{web09} to be a likely
  counterpart -- it is faint at 24\m\ and offset by 7$\arcsec$ -- so it
  is not included in our analysis.}   
\end{deluxetable}

\begin{deluxetable}{lccccc}
\setlength{\tabcolsep}{0.05in} 
\tablehead{
\colhead{Name} &
\colhead{RA} &
\colhead{Dec} &
\colhead{Redshift} &
\colhead{24\m\ Flux} &
\colhead{Int. Time} \\
& & & & $\mu$Jy & minutes }
\tablecaption{IRS LAB Targets}
\startdata    
LAB1\_J2143-4423       & 21h42m27.63s & -44d20m30.3s &  2.38 &  240 &
528 \\ 
LAB6a\_J2143-4423     & 21h42m42.80s & -44d30m18.0s &  2.38 &  540 &
120 \\ 
LAB6b\_J2143-4423    & 21h42m42.65s & -44d30m09.2s &  2.38 &  630 &
120 \\ 
LAB7\_J2143-4423       & 21h42m34.99s & -44d27m08.5s &  2.38 &  290 &
264 \\ 
LAB18\_J1714+5015    & 17h14m11.98s & +50d16m01.5s &  2.39 &  590 & 84
\\ 
LAB1\_J1434+3317      & 14h34m10.98s & +33d17m30.9s &  2.66 &  860 &
52   \\ 
\enddata
\end{deluxetable}

\begin{deluxetable}{lcccccl}
\setlength{\tabcolsep}{0.05in} 
\tabletypesize{\footnotesize}
\tablehead{
\colhead{Name} &
\colhead{6.2\m\ flux} &
\colhead{7.7\m\ flux} &
\colhead{6.2\m\ Lum} &
\colhead{7.7\m\ Lum} &
\colhead{L/C Ratio} &
\colhead{SFR} \\
& 10$^{-15}$ & 10$^{-15}$ &10$^{10}$ L$_{\odot }$ & 10$^{10}$ L$_{\odot }$&  & M$_{\odot }$ yr$^{-1}$ \\
& ergs cm$^{-2}$ s$^{-1}$ & ergs cm$^{-2}$ s$^{-1}$ & & & & } 
\tablecaption{\Lya Blob PAH Characteristics}
\startdata    
LAB1\_J2143-4423           & 0.55\p 0.46  & 2.64\p 1.5  &  0.63 &  3.0 & 1.5\p 0.7 & 420 \\
LAB6a\_J2143-4423         & $<$1.3         &  $<$0.9      &  $<$1.5 & $<$1.3 & $<$0.14& $<$130 \\ 
LAB6b\_J2143-4423         & 3.2\p 0.9      &  11.8\p 2.0 &  3.6  &  13.4& 4.8\p1.4 & 2000 \\
LAB7\_J2143-4423           & 2.23\p 0.7   & 6.9\p 1.6    &  2.6 &  8.0& 6.1\p3.4 & 1200 \\
LAB18\_J1714+5015         & $<$1.2         & $<$0.9       &  $<$1.3  &  $<$1.3 & $<$0.11 & $<$140 \\
LAB1\_J1434+3317           & 2.25\p 0.9   & 9.5\p 2.3     &  3.4  &  14.3 & 0.81\p0.16 & 2200 
\enddata
\end{deluxetable}

\end{document}